\documentclass[preprint,12pt]{elsarticle}

\usepackage{bm}  
\usepackage{amssymb}
\usepackage{amsthm}
\usepackage{amsmath}
\usepackage[linesnumbered,ruled,lined]{algorithm2e}
\graphicspath{{}} 
\usepackage{booktabs}
\usepackage{multirow}
\usepackage{hyperref}
\pdfoutput = 1

\usepackage[outline]{contour}
\newcommand*{\fancy}[1]{{\color{white}\contour{black}{#1}}}

\journal{Elsevier Journal}

\begin{document}

\begin{frontmatter}



\title{An hourglass-free formulation for total Lagrangian smoothed particle hydrodynamics}
\author{Dong Wu}
\ead{dong.wu@tum.de}
\author{Chi Zhang}
\ead{c.zhang@tum.de}
\author{Xiaojing Tang}
\ead{xiaojing.tang@tum.de}
\author{Xiangyu Hu\corref{mycorrespondingauthor}}
\cortext[mycorrespondingauthor]{Corresponding author.}
\ead{xiangyu.hu@tum.de}
\address{TUM School of Engineering and Design,  Technical University of Munich, 85748 Garching, Germany}

\begin{abstract}
The total Lagrangian smoothed particle hydrodynamics (TL-SPH) for elastic solid dynamics 
suffers from hourglass modes which can grow and lead to the failure of simulation for problems with large deformation. 
To address this long-standing issue, 
we present an hourglass-free formulation based on volumetric-devioatric stress decomposition.
Inspired by the fact that the artifact of nonphysical zigzag particle distribution 
induced by the hourglass modes is mainly characterized by shear deformation 
and the standard SPH discretization for the viscous term in the Navier-Stokes (NS) equation, 
the present formulation computes the action of shear stress directly 
through the Laplacian of displacement other than from the divergence of shear stress. 
A comprehensive set of challenging benchmark cases are simulated to demonstrate 
that, while improving accuracy and computational efficiency,
the present formulation
is able to eliminate the hourglass modes 
and achieves very good numerical stability with a single general effective parameter.
In addition, the deformation of a practically relevant stent structure is simulated 
to demonstrate the potential of the present method in the field of biomechanics.
\end{abstract}

\end{frontmatter}


\section{Introduction}\label{sec:introduction}
Smoothed particle hydrodynamics (SPH), 
a fully Lagrangian mesh-free method and originally developed 
for the astrophysical simulation and fluid dynamics 
\cite{lucy1977numerical, gingold1977smoothed}, 
has attracted more and more interest over the past decades 
\cite{randles1996smoothed, luo2021particle, khayyer2022systematic}. 
In SPH method, 
the continuum is represented by particles, 
where the physical properties of the system, e.g. mass and velocity, are located, 
and the discretization of the governing equation is achieved 
through the particle interactions with the help of a Gaussian-like kernel function 
\cite{monaghan2005smoothed, liu2010smoothed, monaghan2012smoothed}. 
Since a significant number of common abstractions, 
which are intrinsically related to numerous physical systems, 
are realized through particle interactions, 
SPH method can be used to discretize the multi-physics equations 
within a unified computational framework \cite{zhang2021sphinxsys},
so that the algorithms, such as neighboring particle search and time stepping, can be shared, 
parallel computation can be simplified 
and efficiency can be greatly improved \cite{sun2021accurate}.
More importantly, 
the unified computational framework permits monolithic and strong coupling, 
which is seamless, spatio-temporal local and conservative locally and globally \cite{matthies2003partitioned, matthies2006algorithms}. 
The fluid-structure interaction (FSI) represents 
a typical multi-physical system in which fluid and solid dynamics are coupled.
Unlike the partition-based coupling for the FSI solver, 
where solid dynamics equations are discretized 
by the finite element method (FEM) \cite{yang2012free, hermange20193d} 
and fluid dynamics equations by SPH method, 
the unified computational framework based on SPH method 
\cite{antoci2007numerical, han2018sph, liu2019smoothed} 
requires that solid dynamics equations, 
particularly those associated with large elastic strain, 
are also discretized by SPH method.

Notwithstanding its promising achievement, 
early attempts indicated that the original SPH method for solid dynamics 
may be unstable and not accurate due to three deficiencies: 
linear inconsistency, tensile instability and hourglass modes. 
The first deficiency is caused by incomplete kernel support 
at domain boundary or with irregular particle distributions \cite{liu2010smoothed}.
As the cure for this problem, 
several algorithms have been proposed in the literature, 
such as the normalized smoothing approach \cite{johnson1996sph}, 
the kernel gradient correction \cite{randles1996smoothed, vignjevic2006sph} 
and the finite particle method (FPM) \cite{liu2006restoring}.
Tensile instability, 
which is characterized by nonphysical fractures  
and void region or particle clustering in solid \cite{swegle1995smoothed} 
and fluid simulations \cite{lind2020review}, respectively,  
arises due to the zero-order inconsistency of the particle discretization 
\cite{rabczuk2004stable}. 
While this deficiency can be generally alleviated by the artificial stress 
\cite{monaghan2000sph, gray2001sph, owen2004tensor} 
and the generalized transport-velocity formulation 
\cite{zhang2017generalized, zhu2021consistency},
it can be completely eliminated by the total Lagrangian SPH (TL-SPH) method, 
in which the kernel function is only evaluated once in the initial undeformed reference configuration \cite{belytschko2000unified} 
unlike the traditional updated Lagrangian SPH (UL-SPH) method, 
without introducing additional correction term 
\cite{bonet2002alternative, vignjevic2006sph}. 
Since its inception,
the TL-SPH method has been successfully applied in many simulations of elastic solid dynamics, 
such as electromagnetically driven rings \cite{de2013total}, 
thermomechanical deformations \cite{ba2018thermomechanical}, 
shell models \cite{maurel2008sph, lin2014efficient, peng2018thick},
FSI \cite{khayyer2018enhanced, liu2019smoothed, zhang2021multi}, 
biomechanics \cite{zhang2021integrative}, etc. 

The artifact of hourglass modes was first observed in FEM simulations,
and is characterized by the zigzag mesh and field pattern
\cite{flanagan1981uniform, jacquotte1984analysis}.
Similar to FEM, 
the hourglass modes in SPH are caused by the deformation gradient remaining unchanged 
when the particles move to the nonphysical zigzag pattern, 
i.e., the zero-energy modes \cite{dyka1997stress, vignjevic2000treatment, vignjevic2009review}.
To address this issue in the UL-SPH method, 
Beissel and Belytschko \cite{beissel1996nodal} 
introduced a stabilization term to the potential energy function
and Vidal et al. \cite{vidal2007stabilized} an artificial viscosity term 
by minimizing a local measure of the Laplacian of the deformation field.
Both schemes have been successfully applied in some benchmark cases,  
however, with empirical case-dependent parameters \cite{o2021fluid}.
A more robust approach is to introduce additional integration or stress points 
between the original particles to present the stress field 
\cite{randles2000normalized, vignjevic2009review}. 
While this approach removes the hourglass modes effectively, 
it increases the complexity of algorithm and computational overhead, 
and suffers from the lack of a rule on determining the location of stress points \cite{islam2019stabilized}.

In the TL-SPH method, 
it is found that introducing artificial viscosity similar to that used in computational fluid dynamics (CFD)
can effectively decrease hourglass modes for the simulation of dynamical problems 
\cite{lee2016new, zhang2022artificial}.
Since these artificial viscosity formulations reply on the particle velocity gradient, 
their validity is questionable when the velocity field becomes flat or less significant.   
More recently, 
Ganzenm{\"u}ller \cite{ganzenmuller2015hourglass} 
introduced an artificial stress method, 
based on the analogy between the SPH and FEM methods, 
to correct the inconsistency due to the zigzag pattern between the local displacement field 
and that linearly predicted from the deformation gradient. 
While effective and computationally efficient,
it may suppresses non-linear part of the displacement field with excessive artificial stiffness \cite{belytschko1983correction, stainier1994improved} 
and, again, requires the empirical case-dependent tuning parameter 
to obtain physically meaningful results \cite{o2021fluid}.

In this paper, 
an hourglass-free formulation without case-dependent tuning parameter 
is developed for the TL-SPH method to simulate elastic solid dynamics. 
Inspired by the fact that the zigzag particle distribution 
is mainly characterized by shear deformation 
and the standard SPH discretization of Laplacian operator for the viscous force 
in the Navier-Stokes (NS) equation \cite{morris1997modeling, monaghan2005smoothed, hu2006multi}, 
we propose a simple and computationally efficient discretization 
for shear deformation and stress based on volumetric 
and devioatric decomposition \cite{simo2006computational}.
The present formulation has been implemented in the TL-SPH method 
with a general effective correction parameter for the error introduced by the kernel summation. 
A set of benchmark cases are first studied to validate the stability, 
accuracy and efficiency of the present formulation. 
Then, a bio-mechanical application,
i.e., the deformation of a stent structure,
is used to demonstrate its potential in the field of bio-mechanics.
The remainder of this paper is organized as follows. 
Section \ref{sec:governingeq} introduces the governing equations of solid dynamics 
together with volumetric and devioatric decomposition.
The details of the present formulation are described in Section \ref{sec:SPHalgorithm}. 
Numerical examples are provided and discussed in Section \ref{sec:examples}, 
and then the concluding remarks are presented in Section \ref{sec:conclusion}. 
For better comparison and future in-depth studies, 
all the computational codes for this study are released in the SPHinXsys repository 
\cite{zhang2020sphinxsys, zhang2021sphinxsys} at \url{https://www.sphinxsys.org}.

\section{Kinematics and governing equations}\label{sec:governingeq}
Considering continuum mechanics in the total Lagrangian framework, 
the kinematics and dynamic equations are expressed 
in terms of the initial, undeformed reference configuration 
$\Omega^0 \subset \mathbb{R}^d$ with $d$ denoting the dimension.
A deformation map $\varphi$ between the initial configuration $\Omega^0$ and 
current deformed configuration $\Omega = \varphi \left( \Omega^0 \right)$ 
describes the body deformation at time $t$ as
\begin{equation}
	\bm{r} = \varphi \left( \bm{r}^0, t \right),
\end{equation}
where $\bm{r}^0$ and $\bm{r}$ are the initial and 
current position of a material point, respectively.
Subsequently, the deformation gradient tensor $\mathbb{F}$ is given by
\begin{equation}\label{deformation-tensor}
	\mathbb{F}  =  \nabla^0 \bm{r} = \nabla^0 \bm{u}  + \mathbb{I},
\end{equation}
 where $\bm{u} = \bm{r} - \bm{r}^0$ is the displacement, 
 $\nabla^0 \equiv \frac{\partial}{\partial \bm{r}^0}$ 
 the gradient operators with respect to the initial configuration 
 $\Omega^0$ and $\mathbb{I} $ the identity matrix.

The conservation equations for mass and momentum in 
the total Lagrangian formulation can be expressed as
\begin{equation}\label{conservation_equation}
	\begin{cases}
		\rho =  J^{-1}\rho^0 \\
		\rho^0 \ddot {\bm{u}} = \nabla^0  \cdot \mathbb{P}^{\operatorname{T}},
	\end{cases}
\end{equation}
where $\rho^0$ and $\rho$ are the initial and current density, respectively, 
$J = \det(\mathbb{F})$, 
$\ddot {\bm{u}}$ the acceleration,
$\mathbb{P}$ the first Piola-Kirchhoff stress tensor, 
and $\operatorname{T}$ the operator of matrix transposition. 
For an ideal elastic or hyperelastic material, 
$\mathbb{P}$ can be given by
\begin{equation}
	\label{first_Piola_kirchhoff_from_S}
	\mathbb{P} = \mathbb{F}\mathbb{S},
\end{equation}
where $\mathbb{S}$ is the second Piola-Kirchhoff stress tensor. 
When the material is liner elastic and isotropic, 
$\mathbb{S}$ can be evaluated via the constitutive equation as 
\begin{equation}\label{linaer_constitutive_relation}
	\begin{split}
		\mathbb{S} &= K \operatorname{tr}\left(\mathbb{E} \right) \mathbb{I} + 2G\left( \mathbb{E} - \frac{1}{3} \operatorname{tr}\left(\mathbb{E} \right) \mathbb{I} \right) \\ 
		&= \lambda\operatorname{tr}\left(\mathbb{E} \right) \mathbb{I} + 2\mu \mathbb{E}, \\ 
	\end{split}
\end{equation}
where $\lambda$ and $\mu$ are Lamé constants, 
$\mathbb{E}  = \frac{1}{2} \left(\mathbb{C} - \mathbb{I}\right)$, 
with $\mathbb{C} = \mathbb{F}^{\operatorname{T}}\mathbb{F}$ 
denoting the right Cauchy deformation tensor, 
is the Green-Lagrangian strain tensor, 
$K = \lambda  + 2\mu /3$ is the bulk modulus 
and $G = \mu$ the shear modulus. 
The relation between the two modulus is given by 
\begin{equation}
	E = 2G \left( 1 + 2\nu \right) = 3K\left( 1 - 2\nu \right),
\end{equation}
where $E$ denotes the Young's modulus and $\nu$ the Poisson ratio. 
To obtain the second Piola-Kirchhoff stress tensor $\mathbb{S}$ 
for a material with nonlinear stress-strain behavior, 
an alternative is to use the strain energy function \cite{ogden1997non},
e.g. for a Neo-Hookean material, defined as follows 
\begin{equation}
	\mathfrak{W}_e = \mu \operatorname{tr}\left(\mathbb{E} \right)  - \mu \ln J + \frac{\lambda}{2} \left(\ln J\right)^2.
\end{equation}
Then, $\mathbb{S}$ can be derived by the partial differentiation of 
the strain energy function as
\begin{equation}
	\mathbb{S} = \frac{\partial \mathfrak{W}_e}{\partial \mathbb{E}} = \mu \mathbb{I} + \left(\lambda \ln J - \mu\right) \mathbb{C}^{-1}.
\end{equation}

The first Piola-Kirchhoff stress tensor $\mathbb{P}$ can also be obtained by the following conversion formula
\begin{equation}
	\label{first_Piola_kirchhoff}
	\mathbb{P} = \fancy{$\tau$}\mathbb{F}^{-\operatorname{T}}, 
\end{equation}
where $\fancy{$\tau$}$ denotes the Kirchhoff stress tensor, 
which can be derived form the following strain energy function 
with volumetric and devioatric decomposition \cite{simo2006computational} as
\begin{equation}
	\mathfrak{W}_e = \mathfrak{W}_v \left( J \right) + \mathfrak{W}_s \left(\bar  {\fancy{$b$}} \right),
\end{equation}
where the volume-preserving left Cauchy-Green deformation gradient tensor 
$\bar  {\fancy{$b$}} = J^ {-\frac{2}{d}}  \fancy{$b$} = \left| \fancy{$b$} \right|^{ - \frac{1}{d}} \fancy{$b$}$ with $\fancy{$b$} = \mathbb{F}\mathbb{F}^{\operatorname{T}}$. 
The volume-dependent strain energy $\mathfrak{W}_v \left( J \right)$ weighted by the bulk modulus $K$ is given by
\begin{equation}
	\mathfrak{W}_v \left( J \right) = \frac{1}{2}K\left[ \frac{1}{2}\left( J^2 - 1 \right) - \ln J \right],
\end{equation}
whereas the shear-dependent strain energy $\mathfrak{W}_s \left(\bar  {\fancy{$b$}} \right)$ weighted by the shear modulus $G$ \cite{yue2015continuum} can be expressed as
\begin{equation}
    \mathfrak{W}_s \left( \bar{ \fancy{$b$}} \right) = \frac{1}{2} G \left( \operatorname{tr} \left( \bar {\fancy{$b$}} \right) - d \right).
\end{equation}
Then, the Kirchhoff stress tensor $\fancy{$\tau$}$ can be derived as
\begin{equation}
	\label{Kirchhoff_stress}
		\fancy{$\tau$} = \frac{\partial \mathfrak{W}_e}{\partial \mathbb{F} } \mathbb{F}^{\operatorname{T}}  
		= \frac{K}{2}\left( J^2 - 1 \right) \mathbb{I} + G \operatorname{dev} \left( \bar {\fancy{$b$}} \right),
\end{equation}
where 
\begin{equation}
	\label{Kirchhoff_stress_part}
	\operatorname{dev} \left( \bar{ \fancy{$b$}} \right)
	= \bar {\fancy{$b$}} -  \frac{1}{d} \operatorname{tr} \left( \bar{ \fancy{$b$}} \right) \mathbb{I} 
	= J^ {-\frac{2}{d}} \left[ \fancy{$b$} - \frac{1}{d} \operatorname{tr} \left( \fancy{$b$} \right) \mathbb{I} \right].
\end{equation}
The deviatoric operator $\operatorname{dev}\left( \bar{ \fancy{$b$}} \right)$ 
returns the trace-free part of $\bar{ \fancy{$b$}}$, 
i.e., $\operatorname{tr} \left( \operatorname{dev}\left( \bar{ \fancy{$b$}} \right) \right)$ is equal to zero. 
In the present formulation, 
we calculate the first Piola-Kirchhoff stress tensor $\mathbb{P}$ from the conversion of the Kirchhoff stress $\fancy{$\tau$}$, i.e. Eq. \eqref{first_Piola_kirchhoff}.

\section{Methodology}\label{sec:SPHalgorithm}
\subsection{Fundamentals of SPH method}
In SPH method, 
the continuum media is discretized by a finite number of Lagrangian particles 
whose interactions are realized by a compact-support kernel function, 
usually a Gaussian-like function, 
to approximate the spatial differential operators. 
Each particle $i$, located at the position vector $\bm{r}_i$ 
and moving with the material velocity $\bm{\dot u}_i$,
carries particle-average field variables such as mass $m_i$, density $\rho_i$, volume $V_i$, etc. 
Then, the discretization for a variable field $f(\bm{r})$ can be written as
\begin{equation}
	f_i  = \int f(\bm{r}) W(\bm{r}_i - \bm{r}, h)d \bm{r},
	\label{particle-average}
\end{equation}
where the kernel function $W(\bm{r}_{i} - \bm{r}, h)$ is radially symmetric with respect to $\bm{r}_i$ 
and $h$ the smoothing length. 
By introducing particle summation, 
this variable field can be approximated by 
\begin{equation}
	f_i \approx \sum\limits_j V_j W(\bm{r}_i - \bm{r}_j, h) f_j = \sum\limits_j \frac{m_j}{\rho_j} W(\bm{r}_i - \bm{r}_j, h) f_j, 
	\label{particle-reconstuction}
\end{equation}
where the summation is conducted over all the neighboring particles $j$ located at the support domain of the particle $i$.

Following Ref. \cite{monaghan2005smoothed}, 
the original SPH approximation of the spatial derivative operator of the variable field $f(\bm{r})$ at particle $i$ can be obtained by 
\begin{equation}
	\label{eq:gradsph}
	\begin{split}
		\nabla f_i & = \int_{\Omega} \nabla f (\bm{r}) W(\bm{r}_i - \bm{r}, h) dV  \\
		& =  - \int_{\Omega} f (\bm{r}) \nabla W(\bm{r}_i - \bm{r}, h) dV \approx  - \sum\limits_j  V_j \nabla_i W_{ij}f_j,
	\end{split}
\end{equation}
where $\nabla_i W_{ij}  = \frac{\partial W\left( r_{ij}, h \right)}{\partial r_{ij}} \bm{e}_{ij} $  
is the derivative of the kernel function 
with $r_{ij}$ denoting the particle distance 
and $\mathbf{e}_{ij}$ the unit vector pointing from particle j to particle i.
Following Ref. \cite{zhang2022review}, we can modify Eq. \eqref{eq:gradsph} into a strong form as
\begin{equation}
	\label{eq:gradsph-strong}
	\nabla f_i = \nabla f_i - f_{i}\nabla 1  \approx   -\sum\limits_j V_j \nabla_i W_{ij} f_{ij}, 
\end{equation}
where $f_{ij} = f_{i} - f_{j}$ is the interparticle difference value. 
This strong-form approximation of the spatial derivative is useful for computing the local structure of a field. 
And Eq. \eqref{eq:gradsph} can also be rewritten into a weak form as
\begin{equation}
	\label{eq:gradsph-weak}
	\nabla f_i = f_{i}\nabla 1 + \nabla f_i \approx   2 \sum\limits_j V_j\nabla W_{ij} \widetilde{f}_{ij},
\end{equation}
where $\widetilde{f}_{ij} = \left(f_{i} + f_{j}\right)/2$ denotes the interparticle average value. 
This weak-form approximation of the derivative is applied to determine the surface integration with respect to a variable 
for solving the conservation law.
Due to the anti-symmetric property of the derivative of the kernel function, 
i.e., $\nabla_i W_{ij} = - \nabla _j W_{ji}$, 
the momentum conservation of the particle system is achieved with Eq. \eqref{eq:gradsph-weak}.

\subsection{Total Lagrangian SPH}
With Eq. \eqref{eq:gradsph-weak} in hand, 
the momentum conservation equation, Eq. \eqref{conservation_equation}, 
is discretized in the TL-SPH formulation as
\begin{equation}\label{discrete_dynamic_equation}
	{\rho_{i}^0}\bm{\ddot u}_i  = \sum\limits_j \left(\mathbb{P}_i {\mathbb{B}_i^0}^{\text{T}} + \mathbb{P}_{j} {\mathbb{B}_j^0}^{\text{T}} \right)  \cdot \nabla_i ^0 W_{ij} V_j^0,
\end{equation}
where $\nabla_i^0 W_{ij}  = \frac{\partial W\left(\bm{r}_{ij}^0, h \right)}{\partial \bm{r}_{ij}^0} \bm{e}_{ij}^0 $
denotes the gradient of the kernel function evaluated at the initial reference configuration. 
Here, we introduce the superscript $\left( \bullet \right)^0$ 
to represent the variable defined at the initial reference configuration. 
The correction matrix $\mathbb{B}^0$ is adopted 
to fulfill first-order completeness as \cite{randles1996smoothed, bonet2002simplified, vignjevic2009review}
\begin{equation}
	\mathbb{B}_i^0  =  \left( 
	\sum\limits_j {V_j^0 \left( {\bm{r}_j^0  - \bm{r}_i^0 } \right) \otimes \nabla_i^0 W_{ij} }
	\right)^{-1}.
\end{equation}
Note that the correction matrix in the TL-SPH formulation is symmetric and computed only once.
The deformation tensor $\mathbb{F}$ is updated by its change rate approximated by using Eq.\eqref{eq:gradsph-strong} as
\begin{equation}
	\label{deformation_tensor_change_rate}
	\frac{d\mathbb{F}_i}{dt}  = \dot{\mathbb{F}}_i = \sum\limits_j V_j^0 \left(\bm{\dot u}_j  - \bm{\dot u}_i \right) \otimes \nabla _i^0 W_{ij} \mathbb{B}_i^0,
\end{equation}

Following Ref. \cite{zhang2022artificial}, 
an artificial damping stress $\fancy{$\tau$}_d$ based on the Kelvin-Voigt type damper is introduced when calculating Kirchhoff stress $\fancy{$\tau$}$ as
\begin{equation}
	\label{Kirchhoff_stress2}
	\fancy{$\tau$}_d = \frac{\gamma}{2}  \cdot \frac{d \fancy{$b$}}{dt},
\end{equation}
where the artificial viscosity factor $\gamma = \rho  c h/2$ with $c  = \sqrt {K/\rho  }$  and the change rate of the left Cauchy-Green deformation gradient tensor
\begin{equation}
	\frac{d \fancy{$b$}}{dt} = \left[\frac{d\mathbb{F}}{dt} \cdot \mathbb{F}^{\operatorname{T}} + \mathbb{F}  \cdot \left( \frac{d\mathbb{F}}{dt} \right)^{\operatorname{T}} \right].
\end{equation}

\subsection{Hourglass-free formulation}\label{sec:hourglass}
\begin{figure}[htb!]
	\centering
	\includegraphics[width=\textwidth]{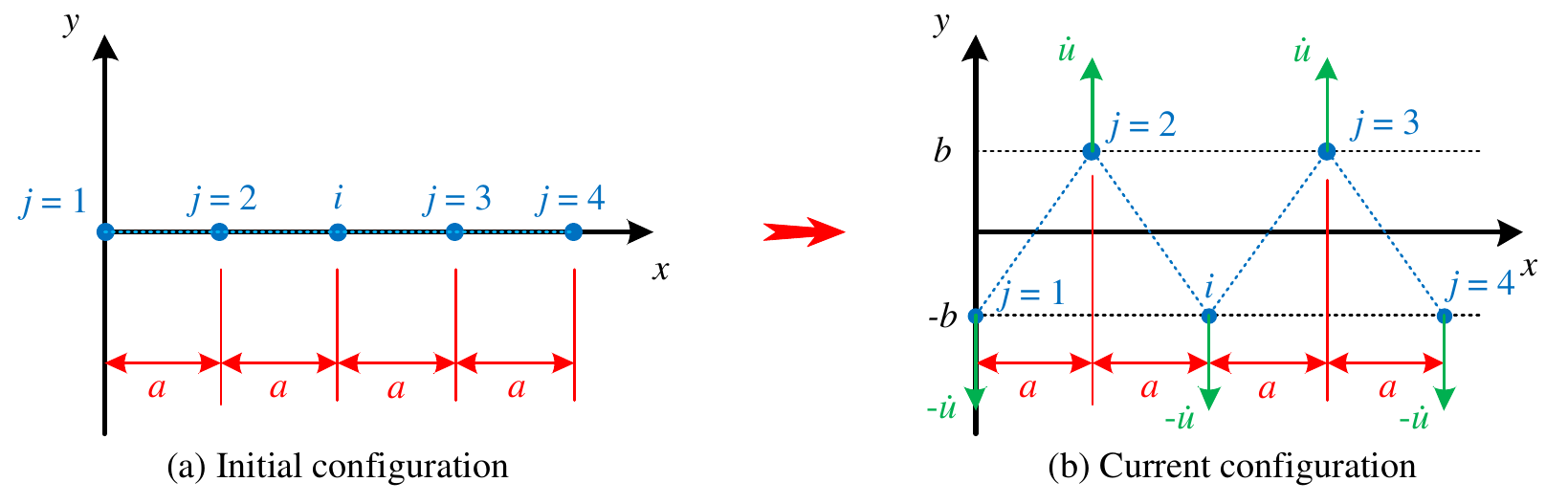}
	\caption{Schematic of zero-energy modes by considering the simple case
		where a single row of particles is uniformly distributed 
		along the $x$-axis in the initial configuration.
		Note that, when applying Eq. (\ref{deformation_tensor_change_rate}),
		the deformation gradient tensor remains vanishing under the action of shearing.}
	\label{figs:zero_energy_modes}
\end{figure}

Although the aforementioned TL-SPH formulation guarantees the first-order consistency 
and avoids the tensile instability, 
the deficiency of hourglass modes still persists 
often when there is large strain or deformation \cite{belytschko2000unified}.
More specifically, the gradient operator in Eq. (\ref{deformation_tensor_change_rate})
averages the relative velocities respected to all neighboring particles,
leads to a smeared-out mean field at the particle center.
This mean approximation may results vanishing 
deformation gradient and thus the stress field
when there is a zigzag particle distribution,
as shown in Figure \ref{figs:zero_energy_modes}, 
which actually indicates very large local, especially shear, deformations. 

Based on the observation that the zigzag particle distribution 
exhibits large shear deformation, 
one may consider a more robust formulation 
in which such shear deformation can be captured directly 
without using Eq. (\ref{deformation_tensor_change_rate}).
On the other hand, one may notice that, 
in a standard SPH formulation of the viscous force 
\cite{morris1997modeling, monaghan2005smoothed, hu2006multi}
in weakly compressible NS equation,
the Laplacian operator is directly discretized with the velocity field 
rather than first computing the shear rate and stress tensors,
and from them computing shear force by applying divergence operation 
\cite{takeda1994numerical}. 
Baring these in mind, 
one may try to find an  hourglass-free formulation in which the shear force is 
obtained by the discretization of the Laplacian operator on the displacement 
(analog to velocity in NS equation) field to capture the shear deformation directly, 
other than from the deformation gradient and second Piola-Kirchhoff stress tensor.
Actually, it is shown later that 
such discretization can be achieved with help of Kirchhoff stress decomposition.
 
We first rewrite the Kirchhoff stress by combining the Eqs. \eqref{Kirchhoff_stress}, 
\eqref{Kirchhoff_stress_part} and (\ref{Kirchhoff_stress2}) as
\begin{equation}
	\label{Kirchhoff_stress4}
	\fancy{$\tau$}= \frac{K}{2}\left( J^2 - 1 \right) \mathbb{I}  
	- \frac{1}{d} J^ {-\frac{2}{d}} G \operatorname{tr} \left( \fancy{$b$} \right) \mathbb{I} 
	+ J^ {-\frac{2}{d}} G \fancy{$b$} 
	+ \fancy{$\tau$}_d,
\end{equation}
where the first term of the right-hand side is the Kirchhoff volumetric stress term, 
the second and third terms together give the Kirchhoff devioatric stress
and the fourth is the numerical damping term. 
Since the second term is the component of the elements on the primary diagonal of the Kirchhoff stress tensor, 
the separated third term, donated as $\fancy{$\tau$}_s$, 
actually contains all the shear stress components. 
For the shear part of the first Piola-Kirchhoff stress $\mathbb{P}_s = \fancy{$\tau$}_s \mathbb{F}^{-\operatorname{T}} 
= J^ {-\frac{2}{d}} G \fancy{$b$} \mathbb{F}^{-\operatorname{T}} = J^ {-\frac{2}{d}} G \mathbb{F}$, 
the particle acceleration induced by the $\fancy{$\tau$}_s$, 
together with Eq. \ref{deformation-tensor} and the weakly-compressible assumption, 
can be derived as
\begin{equation}
	\ddot {\bm{u}}_s = \frac{\nabla^0  \cdot \mathbb{P}_s^{\operatorname{T}}}{\rho^0}
	= \frac{J^ {-\frac{2}{d}} G {\nabla^0 \cdot \mathbb{F}^{\operatorname{T}}}}{\rho^0}
	= G\frac{J^ {-\frac{2}{d}}  {\nabla^0}^2 \bm{r}}{\rho^0},
\end{equation}
where the acceleration due to $\fancy{$\tau$}_s$ is calculated directly from the Laplacian operator of the current position vector.

Inspired by the standard SPH discretization of the viscous term in the NS equation \cite{morris1997modeling}, 
we discretize $\ddot {\bm{u}}_s$ in the total Lagrangian formulation as
\begin{equation}
	\label{shear_acceleration}
	\ddot {\bm{u}}_{si} 
	= \zeta G \sum \limits_j  \left(J_i^{-\frac{2}{d}} + J_j^{-\frac{2}{d}} \right)  \frac {\bm{r}_{ij}}{r_{ij}^0}
	\frac{\partial W\left( r_{ij}^0,h \right)}{\partial r_{ij}^0 }
	\frac { V_j^0}{\rho_i^0},
\end{equation}
where the parameter $\zeta$ is slightly different from unit 
due to the numerical error of kernel summation 
and relevant to the smoothing length $h$ and the choice of kernel function \cite{hu2006angular}. 
It will be shown in the numerical examples that 
$\zeta$ is general effective and remains constant for the simulations in this work. 
Note that Eq. \eqref{shear_acceleration} 
combines a standard SPH first derivative 
with a finite difference approximation of the first derivative 
and precisely preserves the linear momentum \cite{morris1997modeling}. 
Besides the shear stress $\fancy{$\tau$}_s$, the remaining Kirchhoff stress, donated as $\fancy{$\tau$}_r$, is expressed as 
\begin{equation}
	\label{Kirchhoff_stress5}
	\fancy{$\tau$}_r= \frac{K}{2}\left( J^2 - 1 \right) \mathbb{I}  
	- \frac{\zeta}{d} J^ {-\frac{2}{d}} G \operatorname{tr} \left( \fancy{$b$} \right) \mathbb{I}
	+ \fancy{$\tau$}_d.
\end{equation}
Note that the correction factor $\zeta$ is also applied in the second term to fulfill the consistency of the Kirchhoff shear stress.
With the $\mathbb{P}_r = \fancy{$\tau$}_r \mathbb{F}^{-\operatorname{T}}$ in hand,
the acceleration $\bm{\ddot u}_{ri}$ of particle $i$, 
induced by the $\fancy{$\tau$}_{ri}$, 
is calculated by using the Eq. \eqref{discrete_dynamic_equation}. 
Finally, the acceleration of the particle $i$ is given as
\begin{equation}\label{acceleration}
	\bm{\ddot u}_i = \bm{\ddot u}_{ri} + \bm{\ddot u}_{si}.
\end{equation}
\subsection{Time integration scheme}
Following Ref. \cite{zhang2021multi}, 
the position-based Verlet scheme is applied for the time integration.
First, the deformation gradient tensor, density, and particle position are updated to the midpoint as 
\begin{equation}\label{eq:verlet-first-half-solid}
\begin{cases}
\mathbb{F}^{n + \frac{1}{2}} = \mathbb{F}^{n} + \frac{1}{2} \Delta t \dot {\mathbb{F}}^n\\
\rho^{n + \frac{1}{2}} = \rho^0 \frac{1}{J} \\
\bm{r}^{n + \frac{1}{2}} = \bm{r}^n + \frac{1}{2} \Delta t \bm{\dot u}^n.
\end{cases}
\end{equation}
After the calculation of the particle acceleration with Eq. \eqref{acceleration}, the velocity is updated by
\begin{equation}\label{eq:verlet-first-mediate-solid}
\bm{\dot u}^{n + 1} = \bm{\dot u}^{n} +  \Delta t  \bm{\ddot u}.
\end{equation}
Finally, the change rate of deformation gradient tensor $\dot {\mathbb{F}}^{n+1}$ with Eq. \eqref{deformation_tensor_change_rate} is calculated 
and the deformation gradient tensor and position of particles are updated to a new time step with 
\begin{equation}\label{eq:verlet-second-final-solid}
\begin{cases}
\mathbb{F}^{n + 1} = \mathbb{F}^{n + \frac{1}{2}} + \frac{1}{2} \Delta t \dot {\mathbb{F}}^{n+1}\\
\rho^{n + 1} = \rho^0 \frac{1}{J} \\
\bm{r}^{n + 1} = \bm{r}^{n + \frac{1}{2}} + \frac{1}{2} \Delta t \bm{\dot u}^{n + 1}.
\end{cases}
\end{equation}
To maintain the numerical stability, the time step $\Delta t$ is given as
\begin{equation}\label{eq:dt}
	\Delta t   =  \text{CFL} \min\left(\frac{h}{c_v + |\bm{\dot u}|_{max}},
	\sqrt{\frac{h}{|\bm{\ddot u}|_{max}}} \right).
\end{equation} 
Note that the present Courant-Friedrichs-Lewy (CFL) number is set as $0.6$.

\section{Numerical examples}\label{sec:examples}
In this part, 
a set of benchmark tests where analytical or numerical reference data 
in literature are available for qualitative and quantitative comparison
are studied to demonstrate the accuracy and efficiency of 
the present formulation (denoted as TL-SPH-HF).
For comparison, the original formulation in Ref. \cite{han2018sph} 
is denoted as TL-SPH and the artificial stress method in Ref.  \cite{ganzenmuller2015hourglass} TL-SPH-GM.  
Having the validation, 
the deformation of complex stent structures is studied to demonstrate the versatility of the presented formulation. 
The $5th$-order Wendland kernel \cite{wendland1995piecewise} 
with a smoothing length of $h = 1.15~dp$, where $dp$ is the initial particle spacing, and a cut-off radius of $2.3dp$ 
is employed. 
The parameter $\zeta$ is set as 1.07 and remains constant throughout the simulations.

\subsection{Oscillating plate}
\begin{figure}[htb!]
	\centering
	\includegraphics[width=0.6 \textwidth]{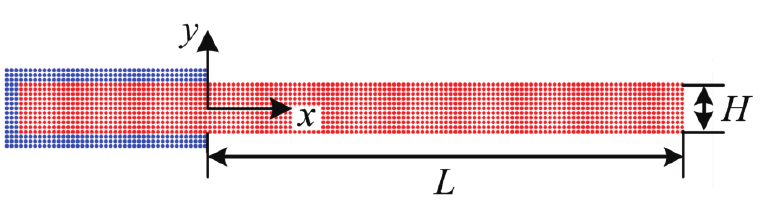}
	\caption{Oscillating plate: Initial configuration.}
	\label{figs:oscillating_plate_setup}
\end{figure}

In this part,
we consider the oscillation of a thin plate with one edge fixed and the others free, 
which has been theoretically \cite{landau1986course} and numerically \cite{gray2001sph, zhang2017generalized} studied in the literature.
This plane strain problem can be modeled by a 2D plate strip of length $L$, 
perpendicular to the fixed edge, 
and thickness $H$. 
Following the Refs. \cite{gray2001sph, zhang2017generalized}, 
the plate strip is clamped between several layers of constrained SPH particles, 
as shown in Figure \ref{figs:oscillating_plate_setup}.
The initial velocity $v_y$, perpendicular to the plate strip, is given by
\begin{equation}
v_y(x) = v_f c_v \frac{f(x)}{f(L)},
\end{equation}
where $v_f$ is a constant that varies with different cases, 
and 
\begin{equation}
	\begin{split}
	f(x) &= \left(\sin(kL) + \sinh(kL) \right) \left(\cos(kx) - \cosh(kx) \right) \\
              & - \left(\cos(kL) + \cosh(kL) \right) \left(\sin(kx) - \sinh(kx) \right)
	\end{split}
\end{equation}
with $k$ determined by 
\begin{equation}
	\cos(kL) \cosh(kL) = -1
\end{equation}
and $kL = 1.875$. 
The material properties are set as follows: 
density $\rho_0 = 1000.0 ~\text{kg} / \text{m}^3$, 
Young’s modulus $E = 2.0~\text{MPa}$ and Poisson’s ratio $\nu$ varies for different cases. 
The frequency $\omega$ of the oscillating plate is theoretically given by 
\begin{equation}
	\omega ^2 = \frac{E H^2 k^4}{12 \rho \left(1 - \nu^2 \right)}.
\end{equation}

\begin{figure}[htb!]
	\centering
	\includegraphics[trim = 2mm 6mm 2mm 2mm, width=\textwidth] {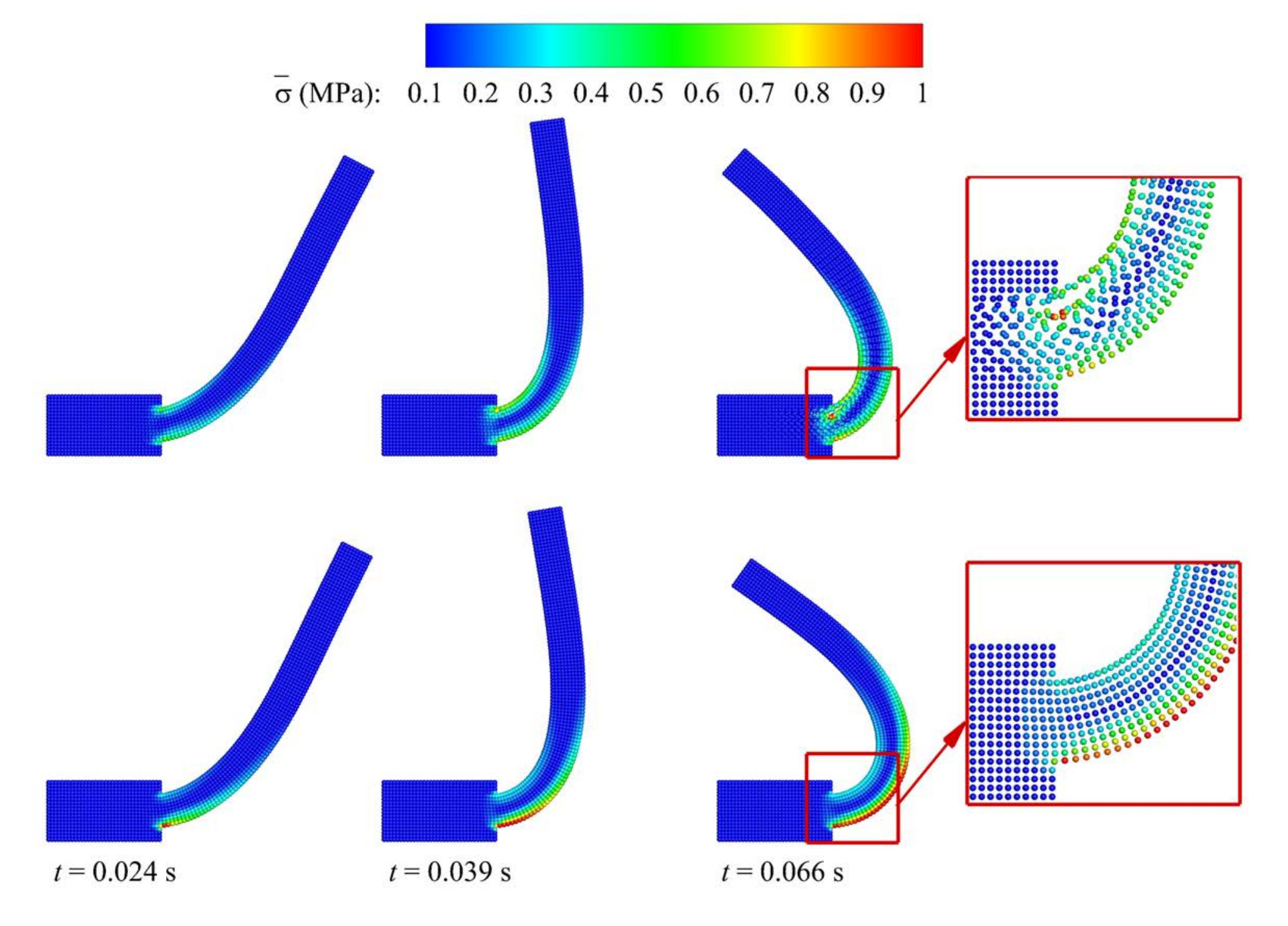}
	\caption{Oscillating plate: Comparison of the deformed configuration colored by von Mises stress $\bar\sigma$ at serial time instants obtained by the TL-SPH (top panel) and the TL-SPH-HF (bottom panel) 
	with the length $L = 0.2~\text{m}$, height $H = 0.02~\text{m}$, $v_f = 0.15$ and spatial particle discretization $H / dp = 10$. 
		The material is modeled with density $\rho_0 = 1000.0 ~\text{kg} / \text{m}^3$, Young’s modulus $E = 2.0~\text{MPa}$, and Poisson’s ratio $\nu = 0.3975$. }
	\label{figs:oscillating_plate_comparison}
\end{figure}

\begin{figure}[htb!]
	\centering
	\includegraphics[trim = 2mm 6mm 2mm 2mm, width=0.5\textwidth] {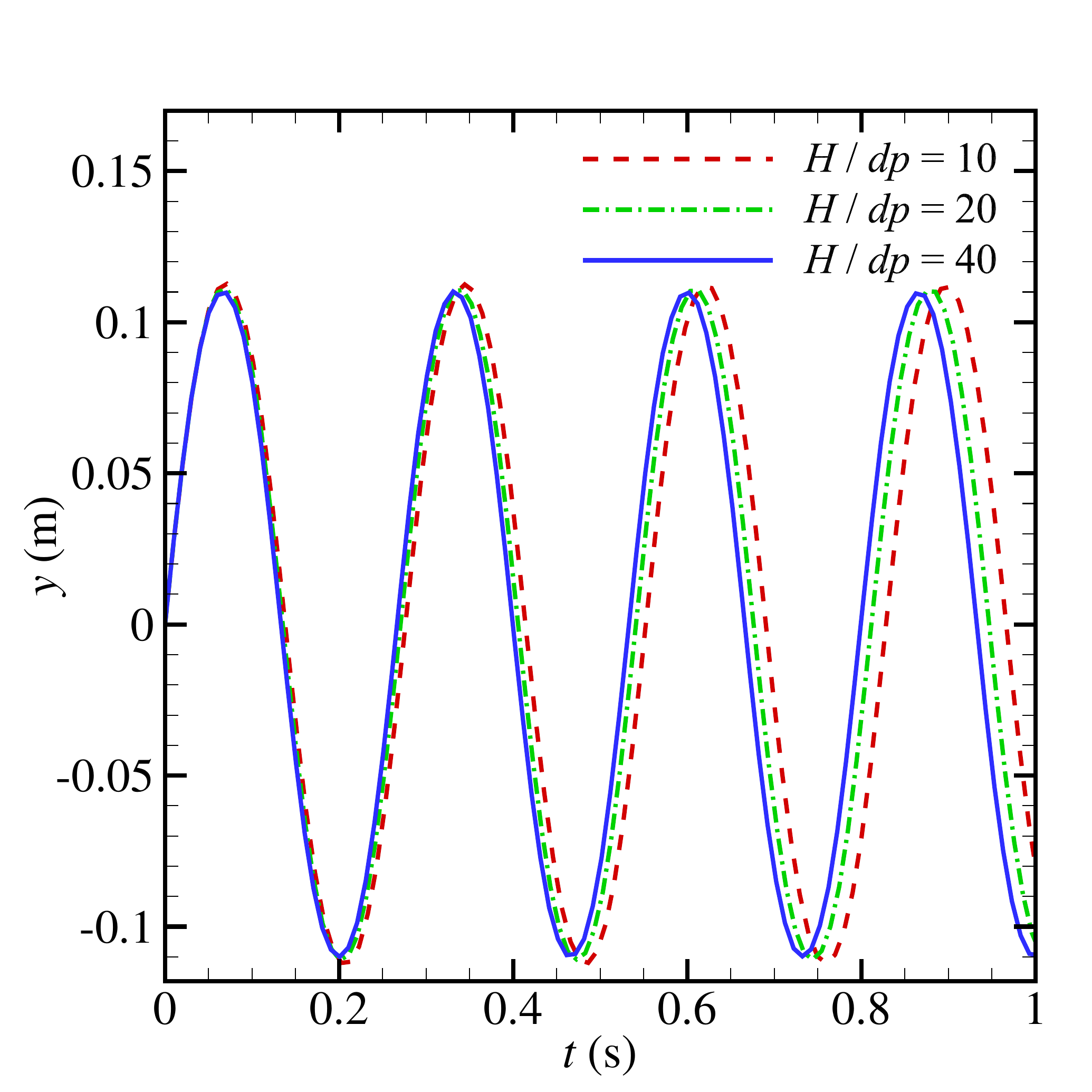}
	\caption{Oscillating plate: Time history of the vertical position $y$ observed at the midpoint of the plate strip end obtained by the TL-SPH-HF 
		with the length $L = 0.2~\text{m}$, height $H = 0.02~\text{m}$, $v_f = 0.05$.
		The material is modeled with density $\rho_0 = 1000.0 ~\text{kg} / \text{m}^3$, Young’s modulus $E = 2.0~\text{MPa}$ and Poisson’s ratio $\nu = 0.3975$. 
	Note that $dp$ is the initial particle spacing.}
	\label{figs:oscillating_plate_convergence}
\end{figure}

Figure \ref{figs:oscillating_plate_comparison} shows the deformed particle configuration with von Mises stress $\bar\sigma$ contour 
obtained by the TL-SPH and TL-SPH-HF for the case of $L = 0.2~\text{m}$, $H = 0.02~\text{m}$, $v_f = 0.15$, $\nu = 0.3975$ 
and the initial particle spacing $dp = H / 10 = 0.002~\text{m}$.
It can be noted that,
while TL-SPH is bale to preserve uniform particle distribution for this problem when the deformation is moderate as in Ref. \cite{han2018sph}, 
its results exhibit particle disorder when the deformation is large,
as shown in Fig. \ref{figs:oscillating_plate_comparison}, 
especially near the region of maximum displacement and stress. 
The larger the deformation of the plate strip is, 
the more pairs of particles stick together, 
which is consistent with that reported in 
Ref. \cite{ganzenmuller2015hourglass} for a static problem 
(see their Figure 6).
On the contrary, 
the TL-SPH-HF, similar to TL-SPH-GM, suppresses such phenomenon
and features smooth deformation and stress fields.
A convergence study and the comparisons between numerical and theoretical solutions are performed to demonstrate the accuracy of the present formulation.
The convergence study tests three different spatial resolutions: $H / dp =10$, $H / dp =20$, and $H / dp =40$. 
Figure \ref{figs:oscillating_plate_convergence} shows 
the vertical position $y$ of the midpoint at the end of the strip 
as a function of time $t$ and initial particle spacing $dp$ when $v_f = 0.05$, 
and exhibits the period and amplitude of the oscillations converge rapidly with increasing resolution.
For quantitative validation, 
Table \ref{tab:oscillating_plate_period1} reports the oscillation period $T$ obtained by the present TL-SPH-HF with the spatial particle resolution $H / dp =40$
and its comparison with the theoretical solution with a wide range of $v_f$ and $\nu$.
The error is about 9.00\% for $\nu = 0.22$ and decreases to about 5.00\% when the Poisson’s ratio is increased to 0.4.
As the thickness is assumed to be very small in the analytical theory, 
Table \ref{tab:oscillating_plate_period2} shows the comparison when the length $L$ remains the same and thickness $H$ is half of its previous value.
A significantly better agreement is obtained with the maximum error decreasing to 2.29\% with $\nu = 0.4$.
It should be noted that when $v_f = 0.15$ and $\nu = 0.4$, the deformation is too large and the plate are in contact with the constrained base, so the period of the plate is not informative.
\begin{table}[htb!]
	\centering
	\caption{Oscillating plate: Quantitative validation of the oscillation period for $L = 0.2~\text{m}$ and $H = 0.02~\text{m}$ with various $v_f $ and $\nu$.}
	\begin{tabular}{ccccc}
		\hline
		$v_f $   & $\nu$  & $T_\text{TL-SPH-HF}$ & $T_\text{Theoretical}$ & Error \\ 
		\hline
		0.01			& 0.22  	& 0.29439     & 0.27009      & 9.00\%\\
		0.05 			& 0.22  	& 0.29428     & 0.27009      & 8.96\%\\
		0.1	 		     & 0.22      & 0.29373     & 0.27009      & 8.75\%\\
		0.15	 		& 0.22      & 0.29374     & 0.27009      & 8.76\%\\
		~\\
		0.01			& 0.30  	& 0.28197     & 0.26412      & 6.76\%\\
		0.05 			& 0.30  	& 0.28166     & 0.26412      & 6.64\%\\
		0.1	 		     & 0.30      & 0.28096     & 0.26412      & 6.38\%\\
		0.15	 		& 0.30      & 0.28126     & 0.26412      & 6.50\%\\
		~\\
		0.01			& 0.40  	& 0.26534     & 0.25376      & 4.56\%\\
		0.05 			& 0.40  	& 0.26473     & 0.25376      & 4.32\%\\
		0.1	 		     & 0.40      & 0.26382     & 0.25376      & 3.96\%\\
		0.15	 		& 0.40      & 0.26656     & 0.25376      & 5.04\%\\
		\hline	
	\end{tabular}
	\label{tab:oscillating_plate_period1}
\end{table}
\begin{table}[htb!]
	\centering
	\caption{Oscillating plate: Quantitative validation of the oscillation period for $L = 0.2~\text{m}$ and $H = 0.01~\text{m}$ with various $v_f $ and $\nu$.}
	\begin{tabular}{ccccc}
		\hline
		$v_f $   & $\nu$  & $T_\text{TL-SPH-HF}$ & $T_\text{Theoretical}$ & Error \\ 
		\hline
		0.01			& 0.22  	& 0.57670     & 0.54018      & 6.76\%\\
		0.05 			& 0.22  	& 0.57205     & 0.54018      & 5.90\%\\
		0.1	 		     & 0.22      & 0.56458     & 0.54018      & 4.52\%\\
		0.15	 		& 0.22      & 0.56677     & 0.54018      & 4.92\%\\
		~\\
		0.01			& 0.30  	& 0.55414     & 0.52824      & 4.90\%\\
		0.05 			& 0.30  	& 0.54638     & 0.52824      & 3.43\%\\
		0.1	 		     & 0.30      & 0.53971     & 0.52824      & 2.17\%\\
		0.15	 		& 0.30      & 0.54027     & 0.52824      & 2.28\%\\
		~\\
		0.01			& 0.40  	& 0.51914     & 0.50752      & 2.29\%\\
		0.05 			& 0.40  	& 0.51074     & 0.50752      & 0.63\%\\
		0.1	 		     & 0.40      & 0.50808     & 0.50752      & 0.11\%\\
		0.15	 		& 0.40      & -     & -      & -\\
		\hline	
	\end{tabular}
	\label{tab:oscillating_plate_period2}
\end{table}

\subsection{Punching strip}\label{sec:punching_strip} 
In this section,
we consider the example of punched rubber where a rubber strip is compressed by punch tools.
This example is a classic challenging test \cite{ganzenmuller2015hourglass} 
not only for meshless methods \cite{chen1996reproducing} 
but also for FEM \cite{chen1996pressure} due to the large deformation.
The rubber strip is defined by a rectangular block of length $L = 9 ~\text{mm}$ and height $H = 3 ~\text{mm}$, 
and its material is modeled with density $\rho_0 = 1100 ~\text{kg} / \text{m}^3$, 
Young’s modulus $E = 1.0~\text{GPa}$ and Poisson’s ratio $\nu = 0.45$.
The punch tools are modeled as rigid rectangular blocks with dimensions 9 mm $\times$ 0.3 mm with the same particle spacing, 
and initialized with a punch velocity of 2 mm/s until the vertical compression ratio of $50\%$ is reached.
A splitting random-choice dynamic relaxation method \cite{zhu2022dynamic} is applied to obtain the quasi-steady solution.

\begin{figure}[htb!]
	\centering
	\includegraphics[trim = 2mm 6mm 2mm 2mm, width=0.8\textwidth]{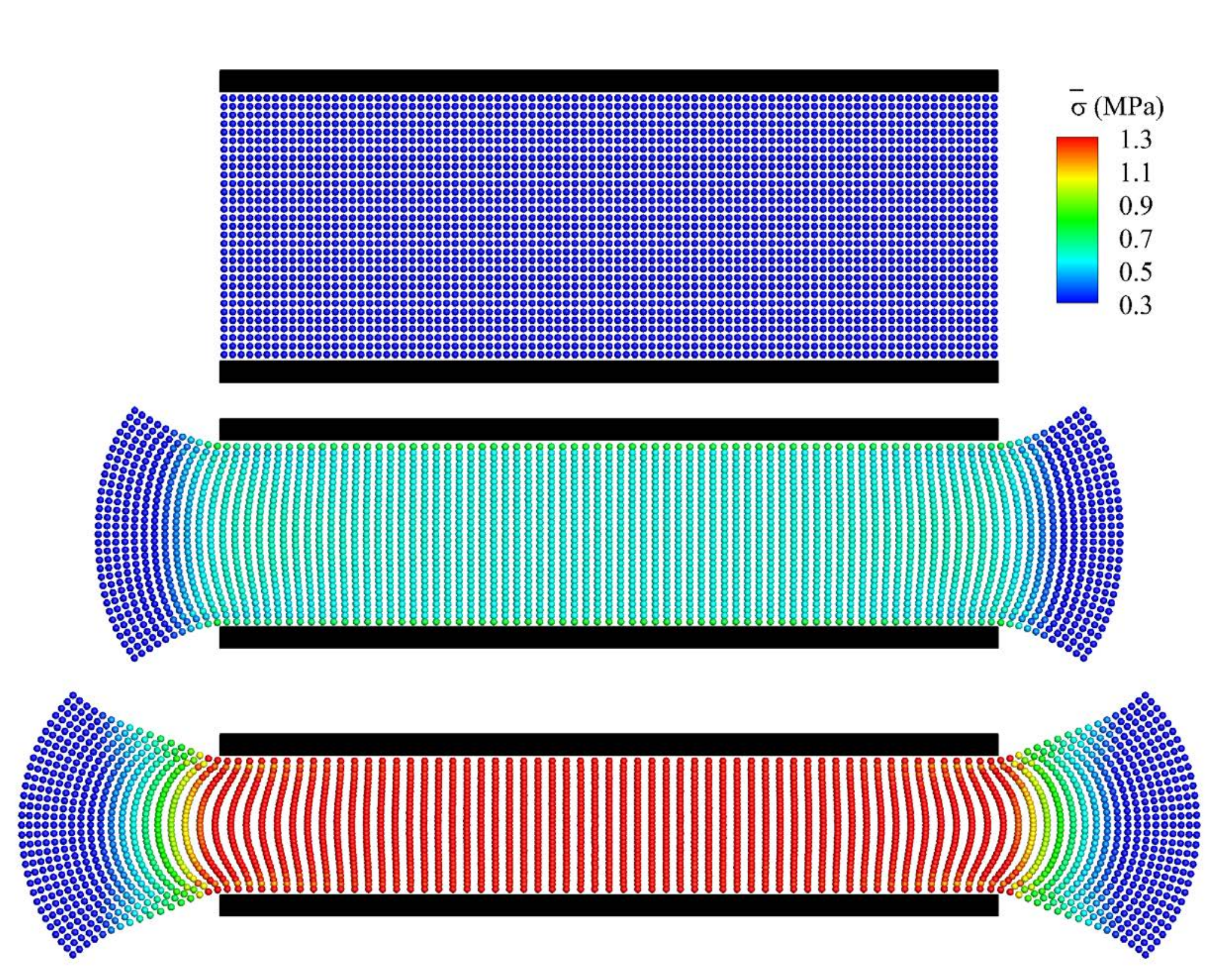}
	\caption{Punching strip: Vertical compression states of 0\%, 25\%, and 50\% with rubber particles colored by von Mises stress $\bar\sigma$.
		The rubber strip is modeled with the length $L = 9~\text{mm}$, height $H = 3~\text{mm}$ and spatial particle discretization $H / dp = 30$, 
		and its material is set as density $\rho_0 = 1100 ~\text{kg} / \text{m}^3$, Young’s modulus $E = 1.0  ~\text{GPa}$ and Poisson’s ratio $\nu = 0.45$.}
	\label{figs:punch_time_points}
\end{figure}

\begin{figure}[htb!]
	\centering
	\includegraphics[trim = 2mm 4mm 2mm 2mm, width=0.8\textwidth]{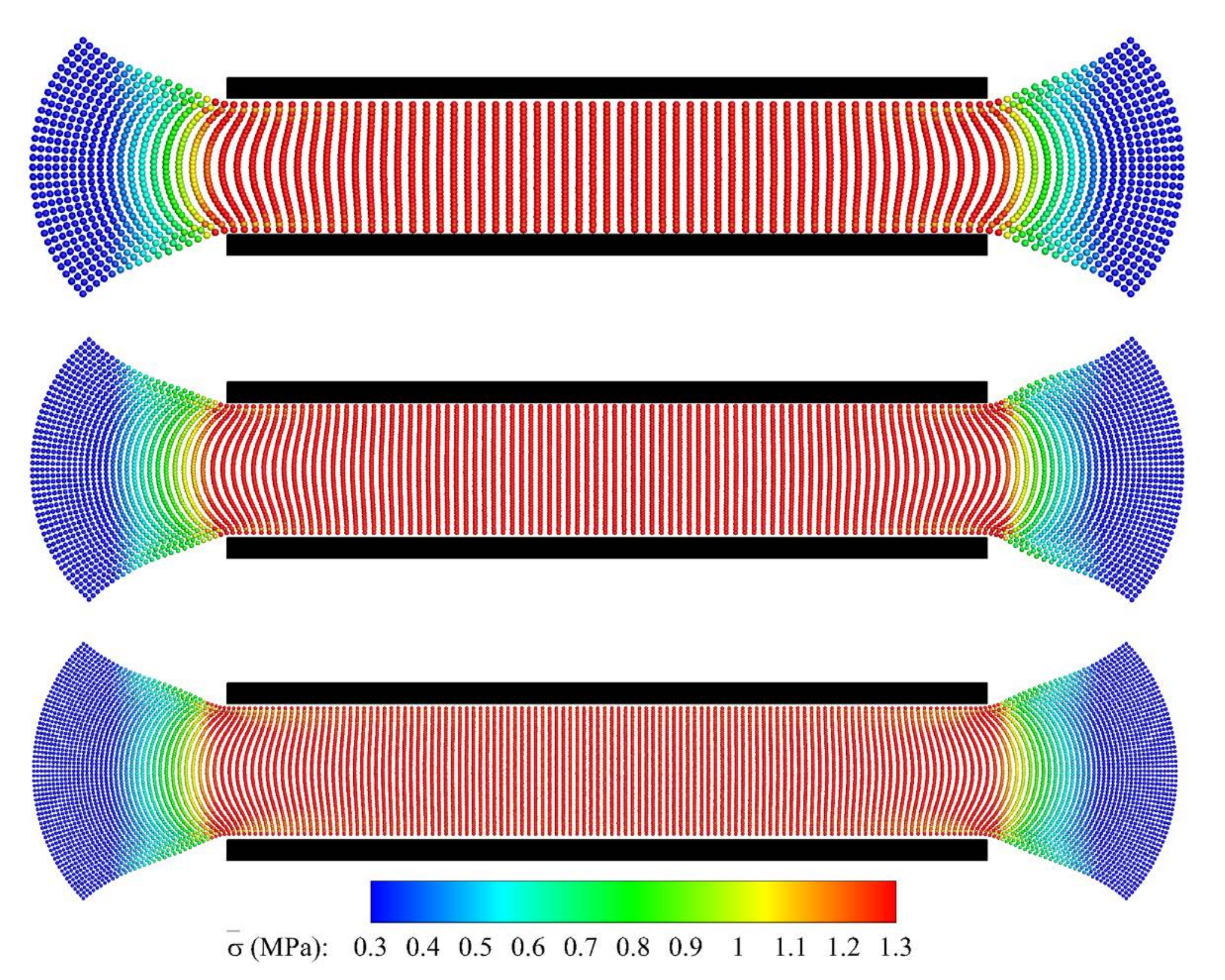}
	\caption{Punching strip: A sequence of particle refinement analysis using the present TL-SPH-HF. 
		Three different spatial resolutions, $H / dp =30$, $H / dp =45$ and $H / dp =60$, are applied. 
		The material is modeled with density $\rho_0 = 1100 ~\text{kg} / \text{m}^3$, Young’s modulus $E = 1.0 ~\text{GPa}$ and Poisson’s ratio $\nu = 0.45$.}
	\label{figs:punch_convergence}
\end{figure}

Figure \ref{figs:punch_time_points} shows the initial and deformed configuration 
colored by von Mises stress $\bar\sigma$ obtained by the present TL-SPH-HF.
As the punch tools compress, 
the rubber strip experiences imposed deformation and the material expands outward towards the open sides.
The very smooth particle distribution and stress field are observed  
even near the sharp corners of the punch tools where the largest deformation exists, 
demonstrating the effectiveness and robustness of the proposed hourglass-free formulation. 
The present deformed configuration is of the volume preservation, 
in contrast to the outcome of TL-SPH-GM stated in Ref. \cite{ganzenmuller2015hourglass} (see their Figure 7), 
which is reflected in the high Poisson's ratios $\nu \in \left[0.45, 0.5\right)$ \cite{greaves2011poisson}.
Figure \ref{figs:punch_convergence} shows the convergence study with particle refinement. 
It can be observed that 
both the deformation pattern and von Mises stress $\bar\sigma$ exhibit good convergence properties.

\subsection{Pulling test}
\begin{figure}[htb!]
	\centering
	\includegraphics[trim = 2mm 6mm 2mm 2mm, width=\textwidth]{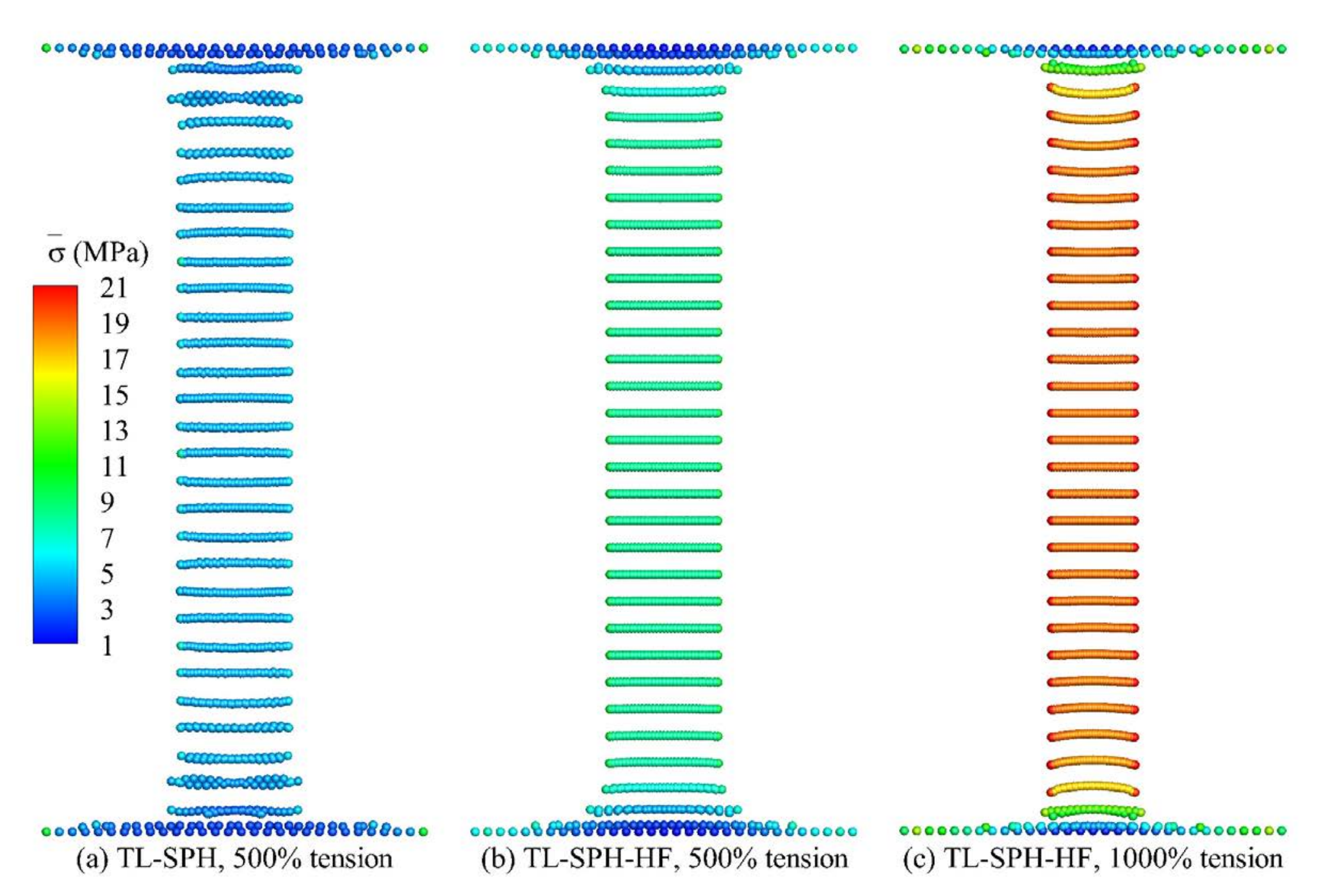}
	\caption{2D pulling rubber square strip: Deformed configuration plotted with von Mises stress $\bar\sigma$ and scaled in the vertical direction for clarity.
		The material parameters are of density $\rho_0 = 1100 ~\text{kg} / \text{m}^3$, Young’s modulus $E = 1.0  ~\text{GPa}$ and Poisson’s ratio $\nu = 0.45$,  
		and the spatial particle discretization is set as $L  / dp = 20$.}
	\label{figs:2D_pulling}
\end{figure}

In this section, 
the 2D pulling rubber strip \cite{chen1996reproducing, ganzenmuller2015hourglass} and 3D pulling rubber cylinder \cite{smith2018stable} are considered 
to investigate the robustness and versatility of the proposed formulation.
Following Ref. \cite{chen1996reproducing, ganzenmuller2015hourglass}, 
the 2D rubber strip is of a square with the side length $L = 2 ~\text{mm}$, 
and its rubber material properties are the same as the previous punching strip test. 
The tensile deformation is initialized by imposing the velocity of  $\bm{v} = \left(0, \pm 0.1 ~\text{mm/s}\right)$ applied to the top and bottom rows of particles respectively.
The initial particle pacing $dp = L /30$ is applied to discretize the system, 
and the splitting random-choice dynamic relaxation method \cite{zhu2022dynamic} is applied to obtain the quasi-static elongation. 
Figure \ref{figs:2D_pulling}(a) and (b) respectively show the particle configuration with von Mises stress contour obtained by the TL-SPH and TL-SPH-HF when the 500\% tension is reached, 
i.e., the length of the strip is increased to $12~\text{mm}$. 
The TL-SPH is unstable and its result exhibits particle disorder in the row of particles, 
and the disorder phenomenon is more obvious near the top and bottom boundaries. 
As expected,
the present TL-SPH-HF formulation is able to stably predict the large tensile deformation.
Different with the unrealistic result of TL-SPH-GM reported in Ref. \cite{ganzenmuller2015hourglass} (see their Figure 8), 
the present deformed configuration is of the typical I-shaped cross section of I-beam, 
which is consistent with that of Ref. \cite{chen1996reproducing} 
(see their Figure 7), 
although slight discrepancy is exhibited near the top and bottom boundaries due to the large stress gradient. 
The robustness of the TL-SPH-HF is further demonstrated for a even more challenging case by increasing the stretch to 1000\% in tension, as shown in Figure \ref{figs:2D_pulling}(c).

\begin{figure}[htb!]
	\centering
	\includegraphics[trim = 2mm 6mm 2mm 2mm, width=\textwidth]{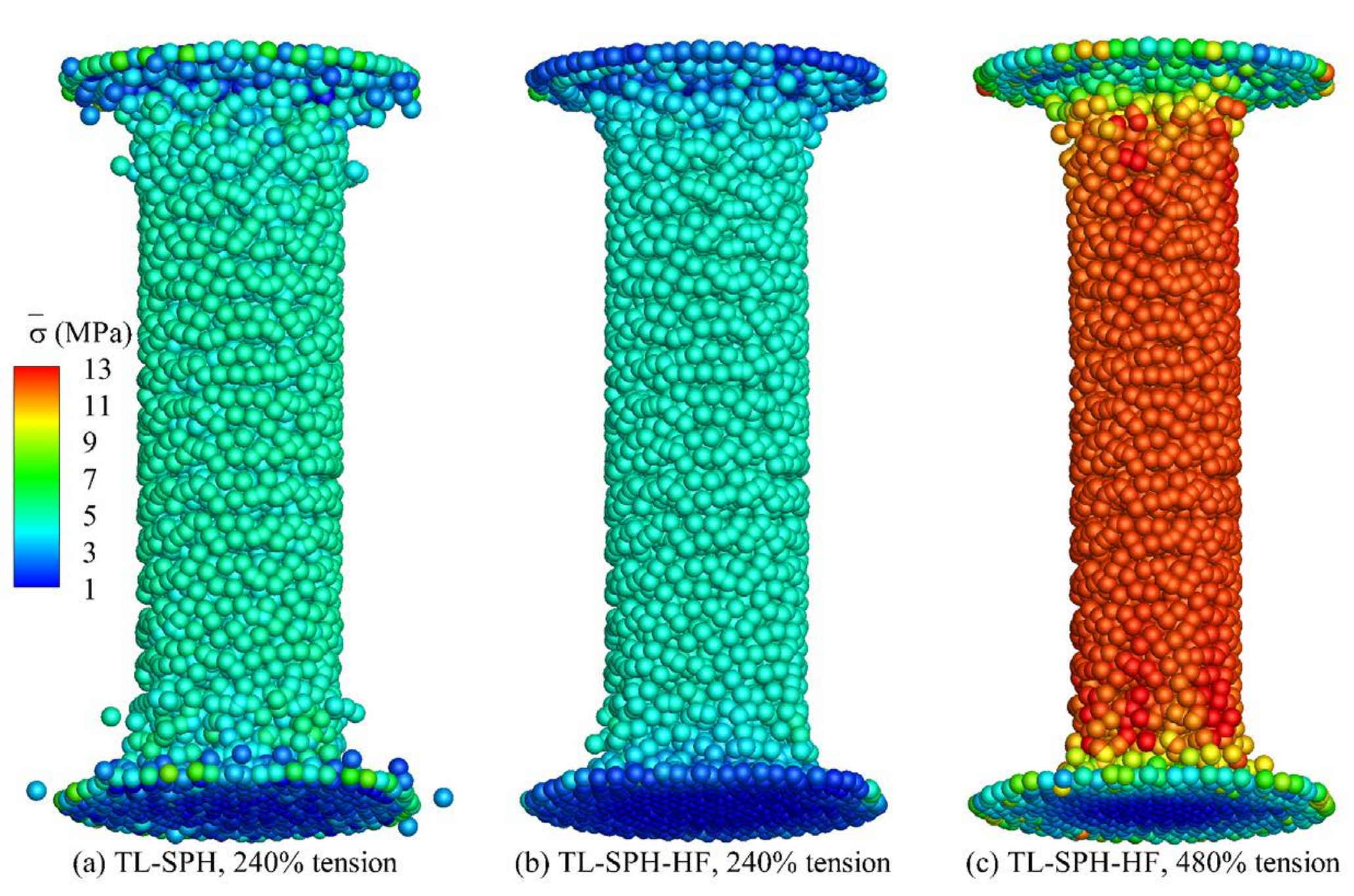}
	\caption{3D pulling rubber cylinder: Deformed configuration plotted with von Mises stress $\bar\sigma$ and scaled in the vertical direction for clarity.
		The material parameters are of density $\rho_0 = 1100 ~\text{kg} / \text{m}^3$, Young’s modulus $E = 1.0  ~\text{GPa}$ and Poisson’s ratio $\nu = 0.49$, 
		and the spatial particle discretization is set as $H / dp = 20 $.}
	\label{figs:3D_pulling}
\end{figure}

The 2D pulling test can be extended to 3D by considering the initial configuration of a cylinder 
with the radius $R = 1~\text{mm}$ and height $H = 2~\text{mm}$.
The Poisson’s ratio is changed to $\nu = 0.49$ following the Ref. \cite{smith2018stable}, 
and the initial particle spacing $dp = 0.1~\text{mm}$.
The body-fitted particle generator \cite{zhu2021cad} is applied to generate initial particle distribution. 
Figure \ref{figs:3D_pulling}(a) and (b) respectively show the deformed configuration colored with von Mises stress obtained by the TL-SPH and TL-SPH-HF when the 240\% tension is reached. 
Again the deformed configuration is of the typical I-shape and is in good agreement with the results from a mesh-based method as in Ref. \cite{smith2018stable}
(see their Figure 5).
Some particles near the top and bottom boundaries run away in the TL-SPH result, 
while the smooth particle and stress distributions are observed in the TL-SPH-HF result. 
Figure \ref{figs:3D_pulling}(c) shows the particle distribution and von Mises stress field for a even more challenging case with 480\% tension. 

\subsection{Bending column}
\begin{figure}[htb!]
	\centering
	\includegraphics{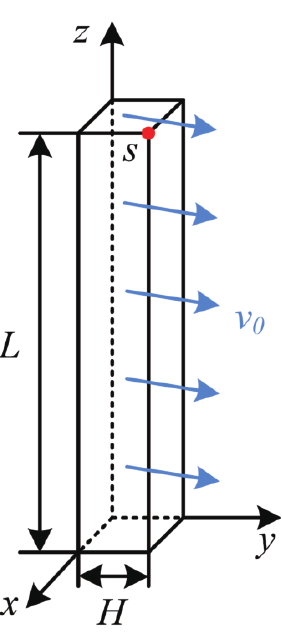}
	\caption{Bending column: Initial configuration.}
	\label{figs:bending_column_setup}
\end{figure}

To further investigate the robustness and accuracy of the present formulation, 
we consider a bending-dominated problem where the numerical solution is available in literature \cite{aguirre2014vertex} for quantitative validation.
Following Ref. \cite{zhang2021integrative}, 
a rubber-like material column spanning the length $L = 6 \operatorname{m}$ and square cross section (height $H = 1 \operatorname{m}$) is clamped on its bottom 
and oscillates freely by imposing an initial uniform velocity $\bm{v_0} = 10\left(\frac{\sqrt{3}}{2}, \frac{1}{2}, 0\right)^{\operatorname{T}} ~\text{m/s}$ as shown in Figure \ref{figs:bending_column_setup}.
The neo-Hookean material model is applied with density $\rho_0 = 1100 ~\text{kg} / \text{m}^3$,  Young’s modulus $E = 17 ~\text{MPa}$ and Poisson’s ratio $\nu = 0.45$.

\begin{figure}[htb!]
	\centering
	\includegraphics[trim = 2mm 4mm 2mm 2mm, width=\textwidth]{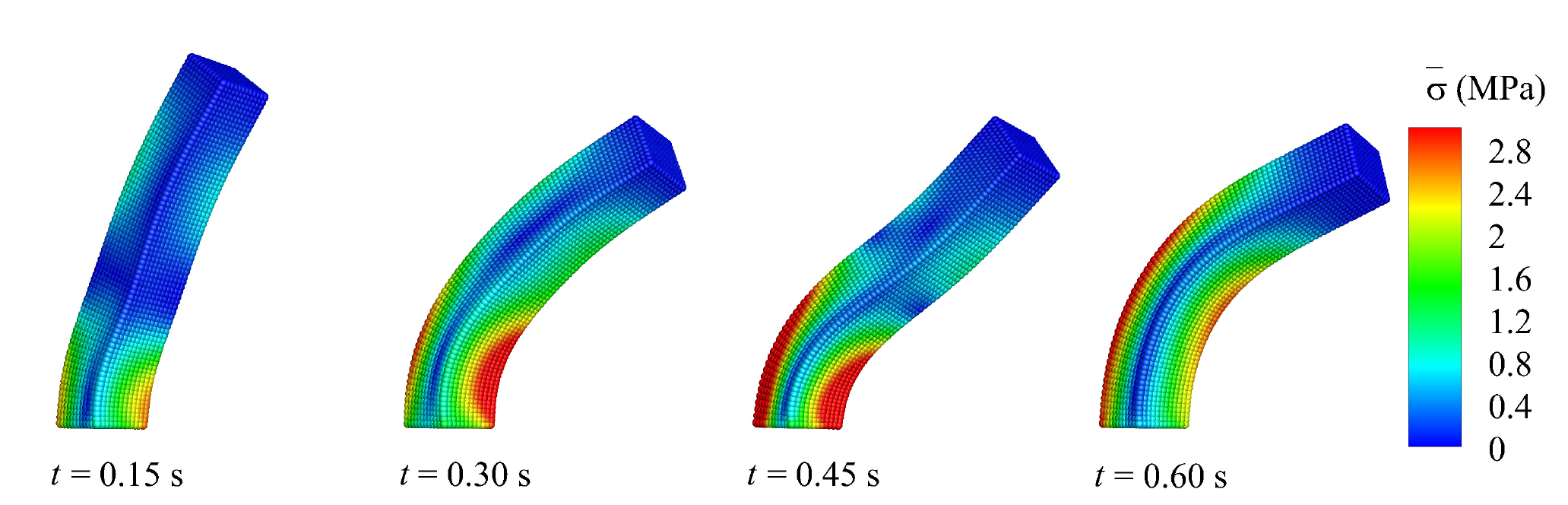}
	\caption{Bending column: Deformed configuration colored by von Mises stress $\bar\sigma$ at serial temporal instants obtained by the present TL-SPH-HF with initial uniform velocity $\bm{v_0} = 10\left(\frac{\sqrt{3}}{2}, \frac{1}{2}, 0\right)^{\operatorname{T}} ~\text{m/s}$. 
		The material is modeled with density $\rho_0 = 1100 ~\text{kg} / \text{m}^3$,  Young’s modulus $E = 17 ~\text{MPa}$  and Poisson’s ratio $\nu = 0.45$, 
		and spatial particle discretization is set as $H / dp =12$ with $H$ denoting the height of the column and $dp$ the initial particle spacing.}
	\label{figs:bending_column_v10}
\end{figure}

\begin{figure}[htb!]
	\centering
	\includegraphics[trim = 0mm 6mm 2mm 2mm, width=\textwidth]{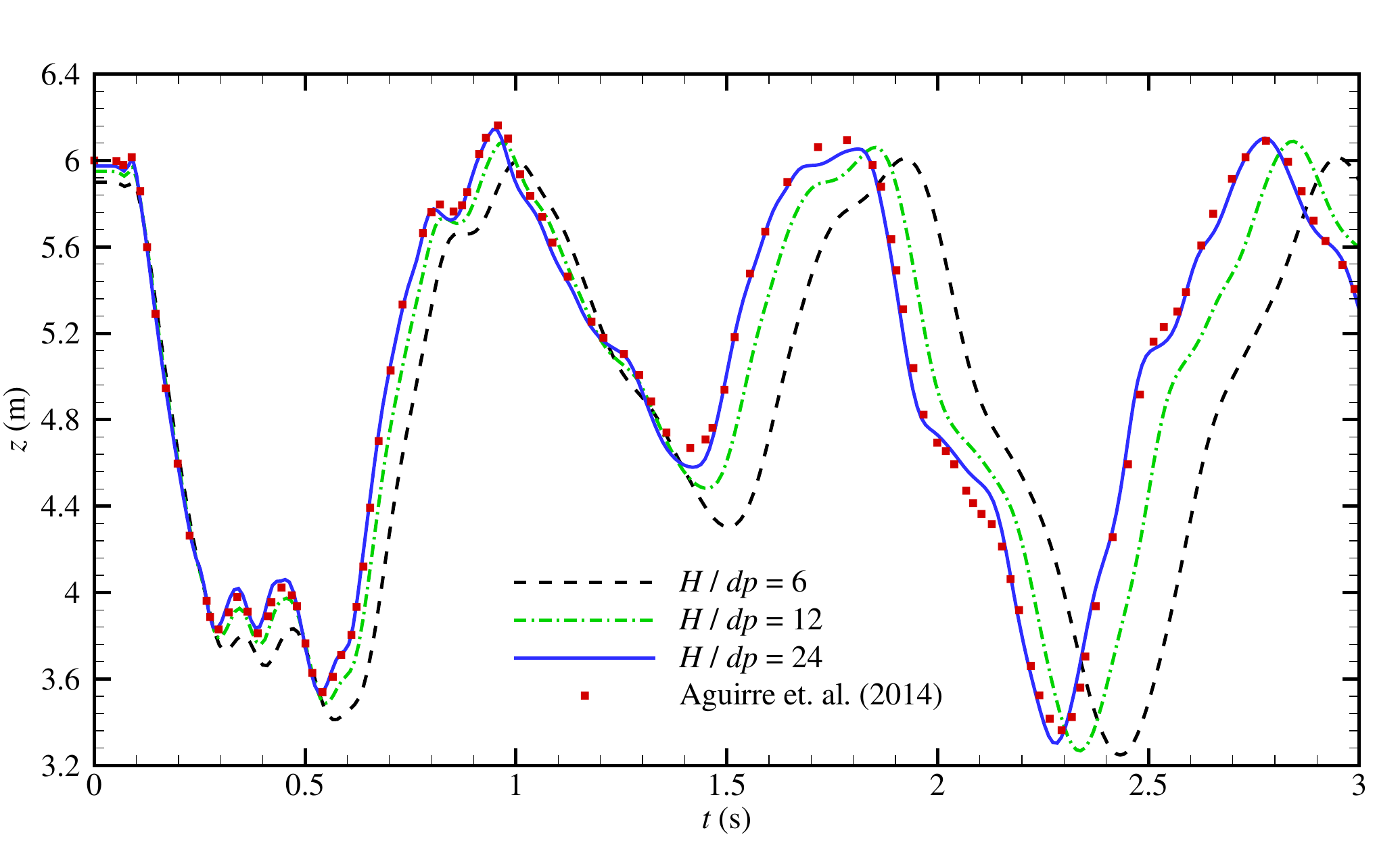}
	\caption{Bending column: Time history of the vertical position $z$ observed at node $S$ obtained by the TL-SPH-HF with three different spatial resolutions and the initial uniform velocity $\bm{v_0} = 10\left(\frac{\sqrt{3}}{2}, \frac{1}{2}, 0\right)^{\operatorname{T}} ~\text{m/s}$,
		and its comparison with that of Aguirre et al. \cite{aguirre2014vertex}. 
		The material is modeled with density $\rho_0 = 1100 ~\text{kg} / \text{m}^3$,  Young’s modulus $E = 17 ~\text{MPa}$, and Poisson’s ratio $\nu = 0.45$.
		Note that $H$ is the height of the column and $dp$ the initial particle spacing.}
	\label{figs:bending_column_convergence}
\end{figure}

\begin{figure}[htb!]
	\centering
	\includegraphics[trim = 2mm 6mm 2mm 2mm, width=\textwidth] {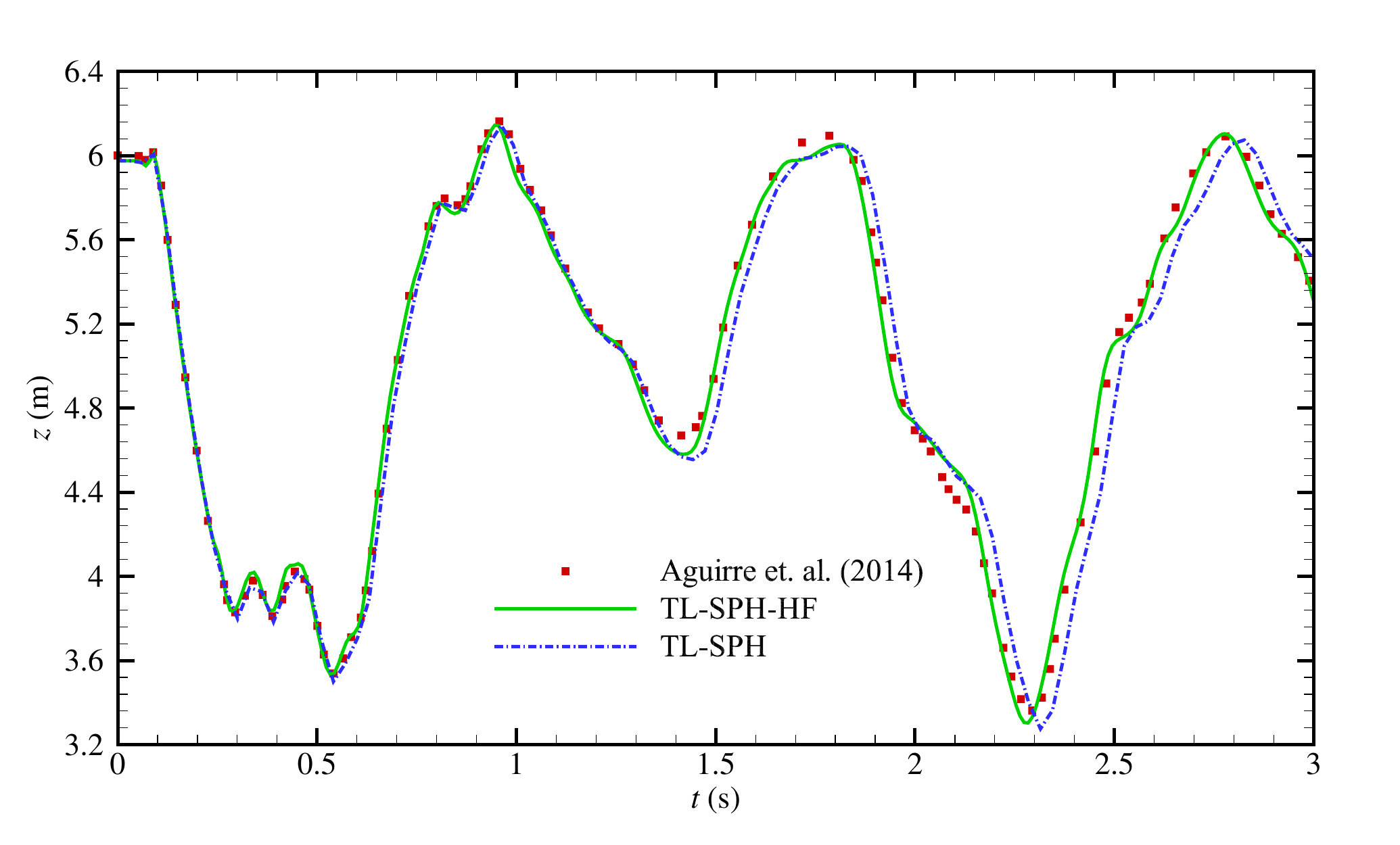}
	\caption{Bending column: Time history of the vertical position $z$ observed at node $S$ obtained by the TL-SPH-HF and TL-SPH with initial uniform velocity $\bm{v_0} = 10\left(\frac{\sqrt{3}}{2}, \frac{1}{2}, 0\right)^{\operatorname{T}} ~\text{m/s}$, 
		and its comparison with that of Aguirre et al. \cite{aguirre2014vertex}. 
		The material is modeled with density $\rho_0 = 1100 ~\text{kg} / \text{m}^3$,  Young’s modulus $E = 17 ~\text{MPa}$ Poisson’s ratio $\nu = 0.45$, 
		and the spatial particle discretization is $H / dp = 24$ with $H$ denoting the height of the column and $dp$ the initial particle spacing.}
	\label{figs:bending_column_comparison}
\end{figure}

Figure \ref{figs:bending_column_v10} shows the time evolution of the deformed configuration colored by von Mises stress contour obtained by the present formulation.
The well-ordered particle distribution and smooth stress field are observed in the present result. 
For quantitative validation,
Figure \ref{figs:bending_column_convergence} reports the time history of the z-axis position of point $S$, 
given in Figure \ref{figs:bending_column_setup},
with different resolutions, 
$H/dp = 6$, $H/dp = 12$, and $H/dp = 24$,
and its comparison with the reference result reported by Aguirre et al. \cite{aguirre2014vertex}.
It can be observed that a good agreement is achieved as the increase of the spatial resolution.
As shown in Figure \ref{figs:bending_column_comparison}, 
compared with the original TL-SPH,
the present TL-SPH-HF shows better agreement with the reference especially in the long run (after $1.5~\text{s}$), 
implying its robustness in the large time scale simulation.

To evaluate the computational performance, 
we analyze the total CPU time of the TL-SPH and TL-SPH-HF for simulating the bending column with physical time of $3~\text{s}$. 
The computations are performed on an Intel Core i7-9700F 3.0GHz 8-core desktop computer. 
Table \ref{tab:computational_efficiency} summarizes the CPU wall-clock time with the corresponding total particle number, 
which shows the cost of calculation is reduced by about 2\% when using the TL-SPH-HF.
\begin{table}[htb!]
	\centering
	\caption{Computational efficiency for the TL-SPH and TL-SPH-HF with different spatial resolutions.}
	\begin{tabular}{ccc}
		\hline
		Model  		 & Resolution  & CPU wall-clock time (s)  \\ 
		\hline
		TL-SPH			& 1,296  	  & 5.10\\
		TL-SPH-HF	& 1,296  	  & 4.98\\
		~\\
		TL-SPH	 		 & 10,368      & 103.89\\
		TL-SPH-HF	 & 10,368      & 99.74\\
		~\\
		TL-SPH			 & 82,944	  & 1777.46\\
		TL-SPH-HF 	 & 82,944	  & 1746.83\\
		\hline	
	\end{tabular}
	\label{tab:computational_efficiency}
\end{table}

\begin{figure}[htb!]
	\centering
	\includegraphics[trim = 0mm 6mm 6mm 4mm, width=\textwidth]{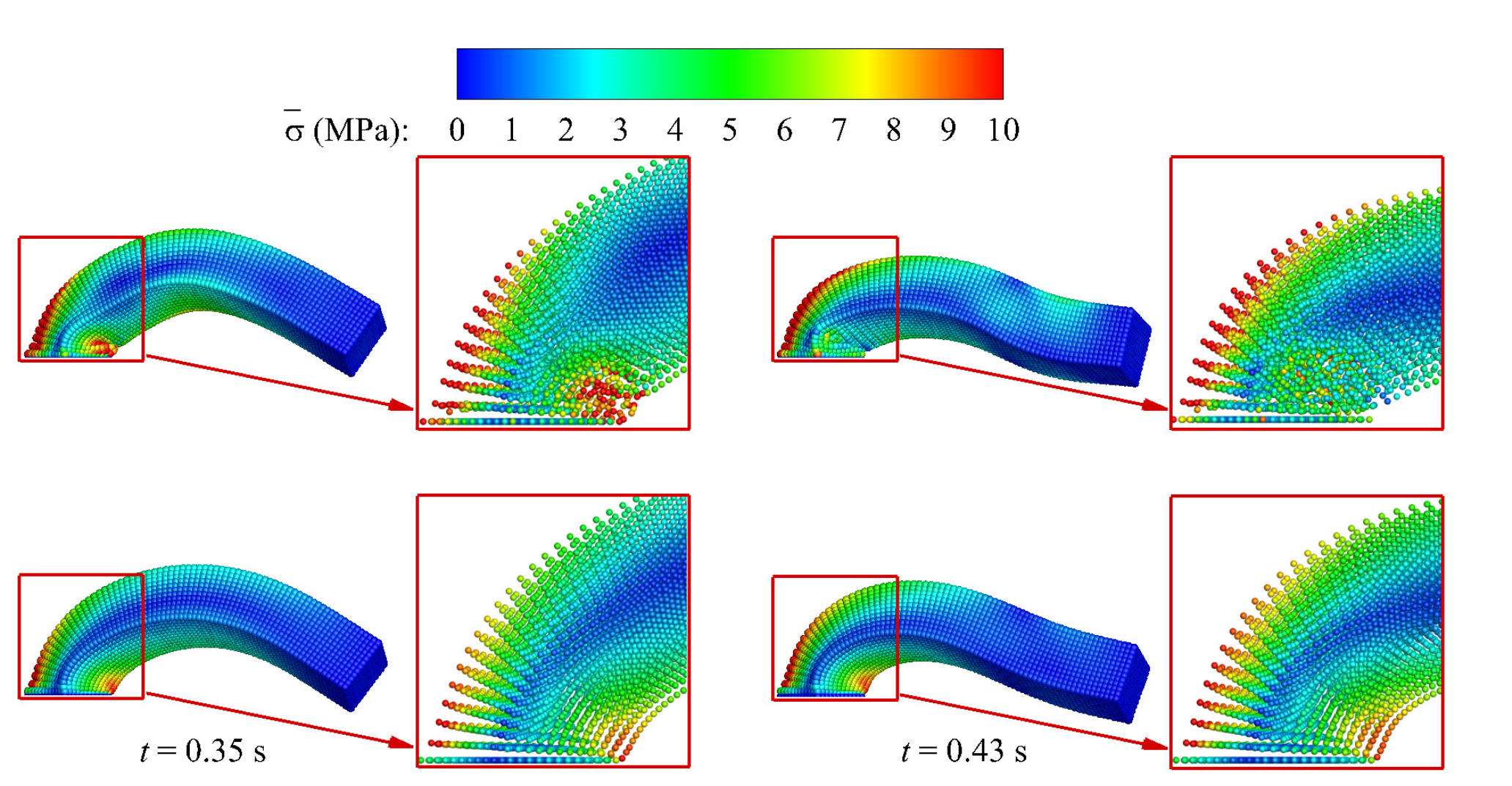}
	\caption{Bending column:  Deformed configuration colored by von Mises stress at two temporal instants obtained by the TL-SPH (top panel) and TL-SPH-HF (bottom panel) with initial uniform velocity $\bm{v_0} = 20\left(\frac{\sqrt{3}}{2}, \frac{1}{2}, 0\right)^{\operatorname{T}} ~\text{m/s}$. 
		The material is modeled with density $\rho_0 = 1100 ~\text{kg} / \text{m}^3$,  Young’s modulus $E = 17 ~\text{MPa}$ and Poisson’s ratio $\nu = 0.45$.
		and the spatial particle discretization is $H / dp = 12$ with $H$ denoting the height of the column and $dp$ the initial particle spacing.}
	\label{figs:bending_column_v20}
\end{figure}

A more challenging problem is studied to show the outperformance of the present formulation by increasing the initial velocity to $\bm{v_0} = 20\left(\frac{\sqrt{3}}{2}, \frac{1}{2}, 0\right)^{\operatorname{T}} ~\text{m/s}$.
As shown in Figure \ref{figs:bending_column_v20}, 
the simulation result of the TL-SPH exhibits noticeable particle disorder, especially near the clamped bottom where the maximum von Mises stress occurs,
while the present TL-SPH-HF captures the very regular particle distribution and smoother stress field,
further demonstrating the robustness of the proposed hourglass-free formulation.

\subsection{Twisting column}
\begin{figure}[htb!]
	\centering
	\includegraphics{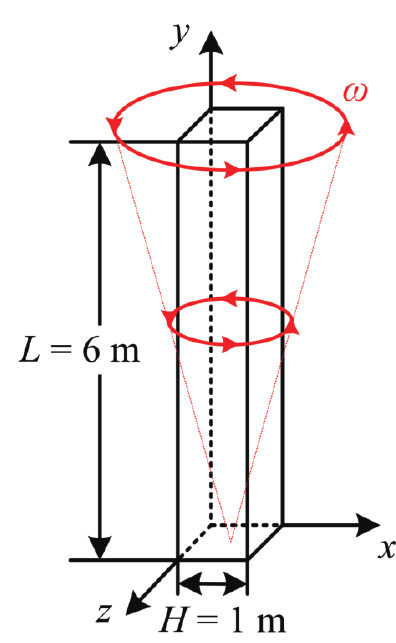}
	\caption{Twisting column: Initial configuration.}
	\label{figs:twsting_column_setup}
\end{figure}

In this section, 
the bending column is extended to a twisting column in line with Refs. \cite{lee2016new, lee2019total, zhang2022artificial}.
As shown in Figure \ref{figs:twsting_column_setup},
the twisting is initialized with a sinusoidal rotational velocity field of $\bm{\omega} = \left[ 0, \Omega_0 \operatorname{sin}\left( \pi y_0 /2 L \right),  0\right]$ with $\Omega_0 = 105~\operatorname{rad/s}$.
Th column is considered as being nearly incompressible with neo-Hookean material,
modeled of density $\rho_0 = 1100 ~\text{kg} / \text{m}^3$, 
Young’s modulus $E = 17 ~\text{MPa}$ and Poisson’s ratio $\nu = 0.499$.

\begin{figure}[htb!]
	\centering
	\includegraphics[trim = 2mm 6mm 2mm 2mm, width=\textwidth]{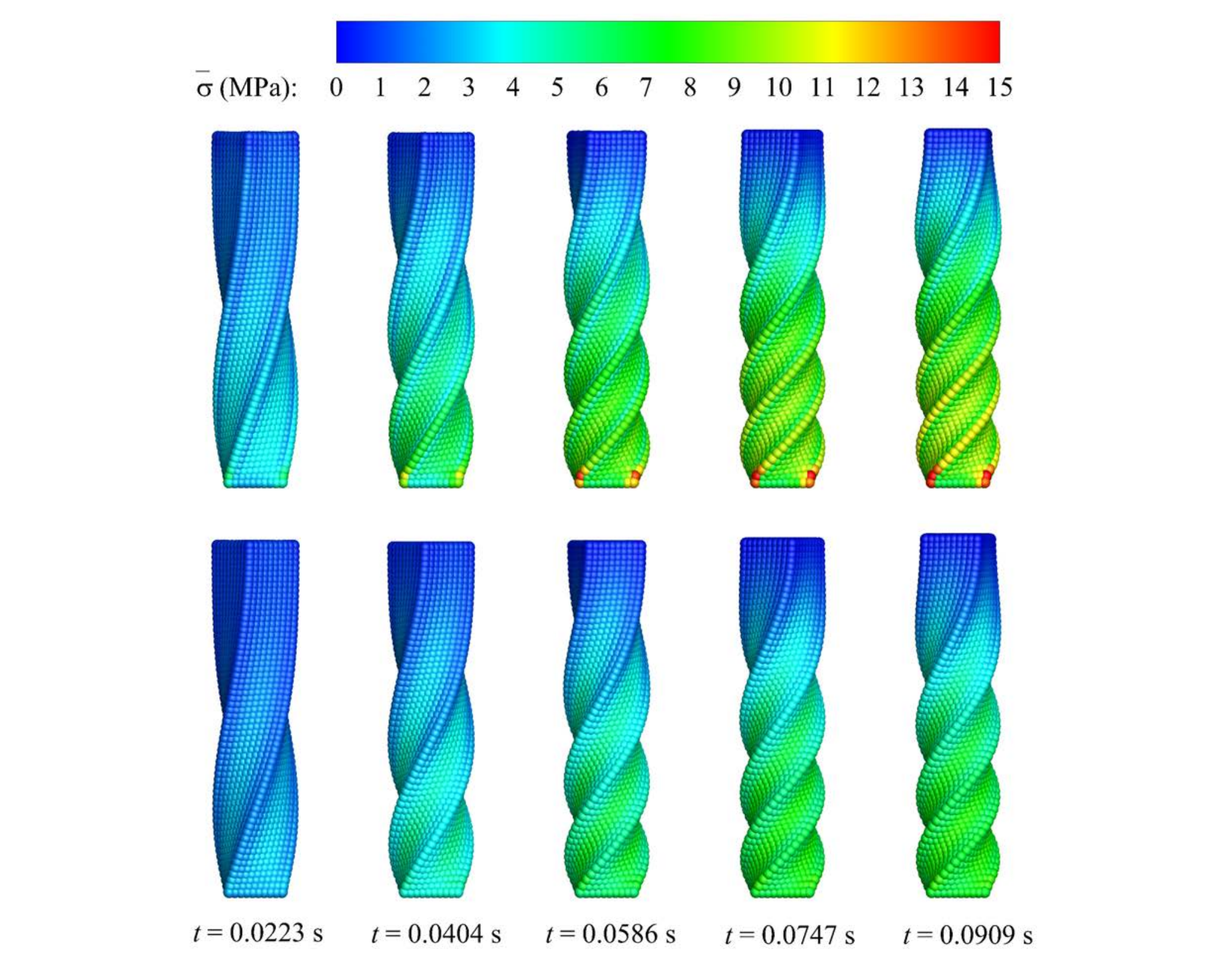}
	\caption{Twisting column: Deformed configuration colored by von Mises stress at different time instants for the TL-SPH (top panel) and TL-SPH-HF (bottom panel) with initial rotational velocity $\bm{\omega} = \left[ 0, \Omega_0 \operatorname{sin}\left( \pi y_0 / 2 L \right),  0\right]$ with $\Omega_0 = 105~\operatorname{rad/s}$. 
		The material is modeled with density $\rho_0 = 1100 ~\text{kg} / \text{m}^3$,  Young’s modulus $E = 17 ~\text{MPa}$ and Poisson’s ratio $\nu = 0.499$, 
		and the spatial particle discretization is set as $H / dp = 10$ with $H$ denoting the height of the column and $dp$ the initial particle spacing.}
	\label{figs:twsting_column_comparison_w105}
\end{figure}

\begin{figure}[htb!]
	\centering
	\includegraphics[trim = 2mm 12mm 2mm 2mm, width=\textwidth]{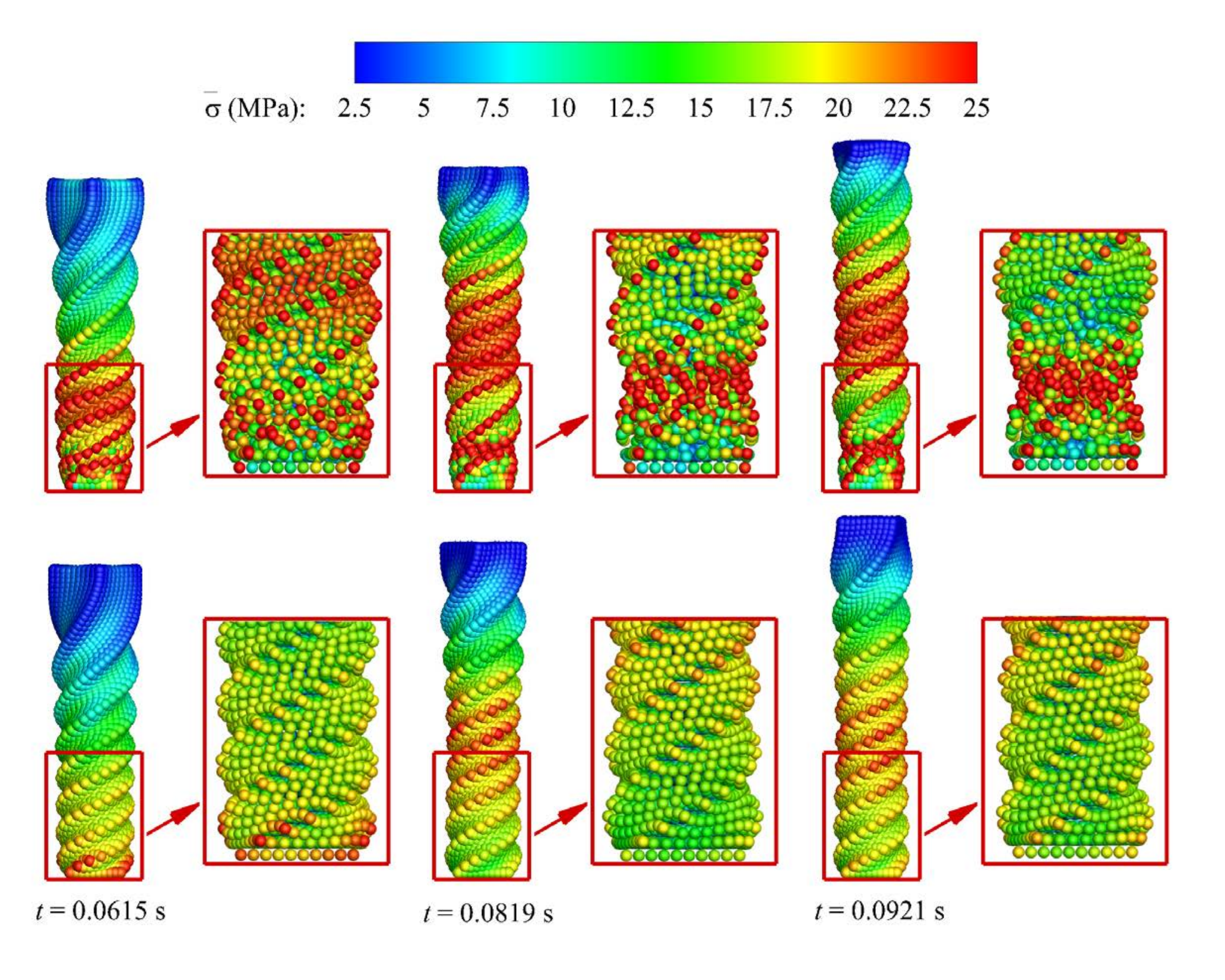}
	\caption{Twisting column: Deformed configuration colored by von Mises stress at different time instants for the TL-SPH (top panel) and TL-SPH-HF (bottom panel) with initial sinusoidal rotational velocity $\Omega_0 = 300~\operatorname{rad/s}$. The material is modeled with density $\rho_0 = 1100 ~\text{kg} / \text{m}^3$,  Young’s modulus $E = 17~\text{MPa}$ and Poisson’s ratio $\nu = 0.49$, and the spatial particle discretization is set as $H / dp = 10$ with $H$ denoting the height of the column and $dp$ the initial particle spacing.}
	\label{figs:twsting_column_comparison_w300}
\end{figure}

\begin{figure}[htb!]
	\centering
	\includegraphics[trim = 4mm 8mm 4mm 4mm, width=0.5\textwidth]{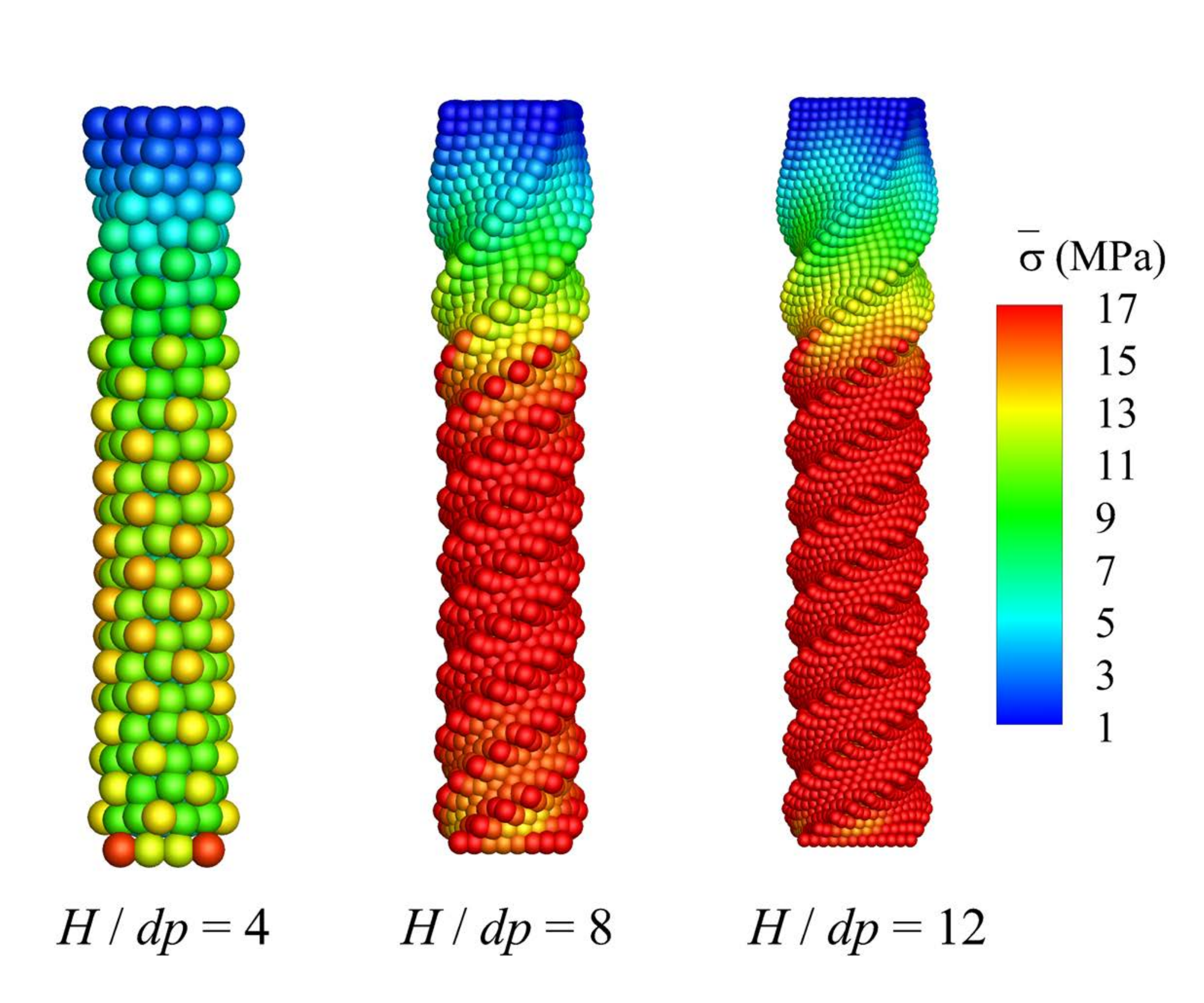}
	\caption{Twisting column: A sequence of particle refinement analysis using the TL-SPH-HF. Results obtained with initial sinusoidal rotational velocity $\Omega_0 = 300~\operatorname{rad/s}$. 
		The material is modeled with density $\rho_0 = 1100 ~\text{kg} / \text{m}^3$,  Young’s modulus $E = 17~\text{MPa}$ and Poisson’s ratio $\nu = 0.49$. 
	    Note that $H$ is the height of the column and $dp$ the initial particle spacing.}
	\label{figs:twsting_column_convergence_w300}
\end{figure}

Figure \ref{figs:twsting_column_comparison_w105} shows the deformed configuration at different time instants with von Mises stress contour obtained by the TL-SPH and TL-SPH-HF.
Both simulations perform well 
and produce very similar results in terms of deformation patterns compared with those in the literature (see Figure 28 in Ref. \cite{lee2016new}),
except small fluctuation of stress near the bottom constrained surface produced by TL-SPH.
A significantly more challenging problem is studied by increasing the initial angular velocity to $\Omega_0 = 300~\operatorname{rad/s}$ with $\nu = 0.49$.
As shown in Figure \ref{figs:twsting_column_comparison_w300}, 
a stable simulation by applying the hourglass-free formulation is demonstrated.
The unstabilized results of the TL-SPH show obvious particle disorder, especially between the second and third spiral lines from the bottom.
On the contrary,
the results calculated by the TL-SPH-HF exhibit the very ordered particle distribution and smooth stress field.
A convergence study is also carried out by sequentially refining the spatial resolution from $H/dp = 4$ to $H/dp = 8$ and $H/dp = 12$.
As shown in Figure \ref{figs:twsting_column_convergence_w300},
both the deformation and von Mises stress $\bar\sigma$ exhibit good convergence properties.

\begin{figure}[htb!]
	\centering
	\includegraphics[trim = 0mm 4mm 0mm 0mm, width=\textwidth]{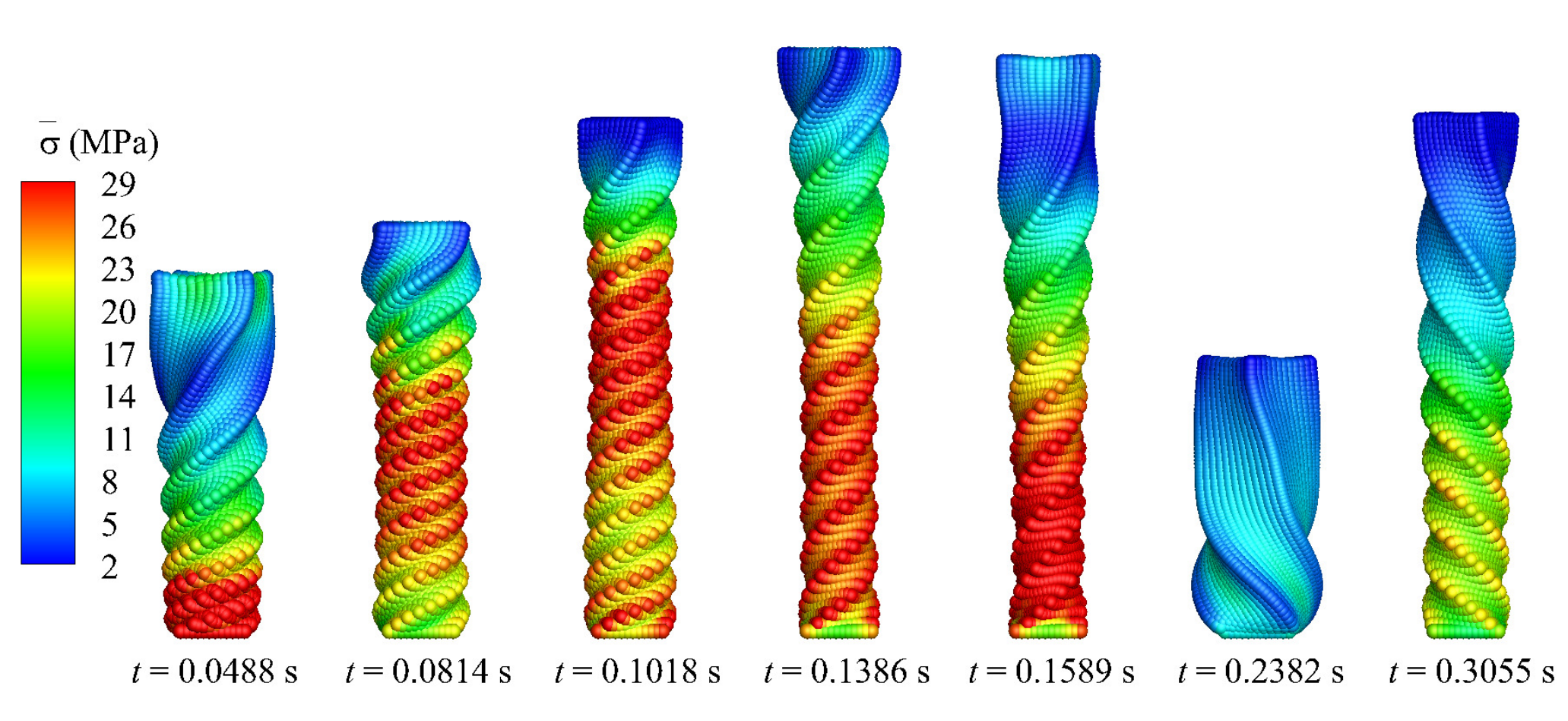}
	\caption{Twisting column: Deformed configuration plotted with von Mises stress at serial time instants obtained by the TL-SPH-HF with initial sinusoidal rotational velocity $\Omega_0 = 400~\operatorname{rad/s}$. 
		The material is modeled with density $\rho_0 = 1100 ~\text{kg} / \text{m}^3$,  Young’s modulus $E = 17~\text{MPa}$ and Poisson’s ratio $\nu = 0.49$,  
		and the spatial particle discretization is set as $H / dp = 12$ with $H$ denoting the height of the column and $dp$ the initial particle spacing.}
	\label{figs:twsting_column_w400}
\end{figure}

Finally, the robustness of the present formulation is further examined by increasing the initial angular velocity to $\Omega_0 = 400~\operatorname{rad/s}$.
Figure \ref{figs:twsting_column_w400} shows the deformed configuration with different time instants.
The extremely large deformations of the whole twisting process, including the recovery process and reverse rotation, are well captured as expected.

\subsection{Stent structure}
\begin{figure}[htb!]
	\centering
	\includegraphics[trim = 0mm 4mm 0mm 0mm, width=\textwidth]{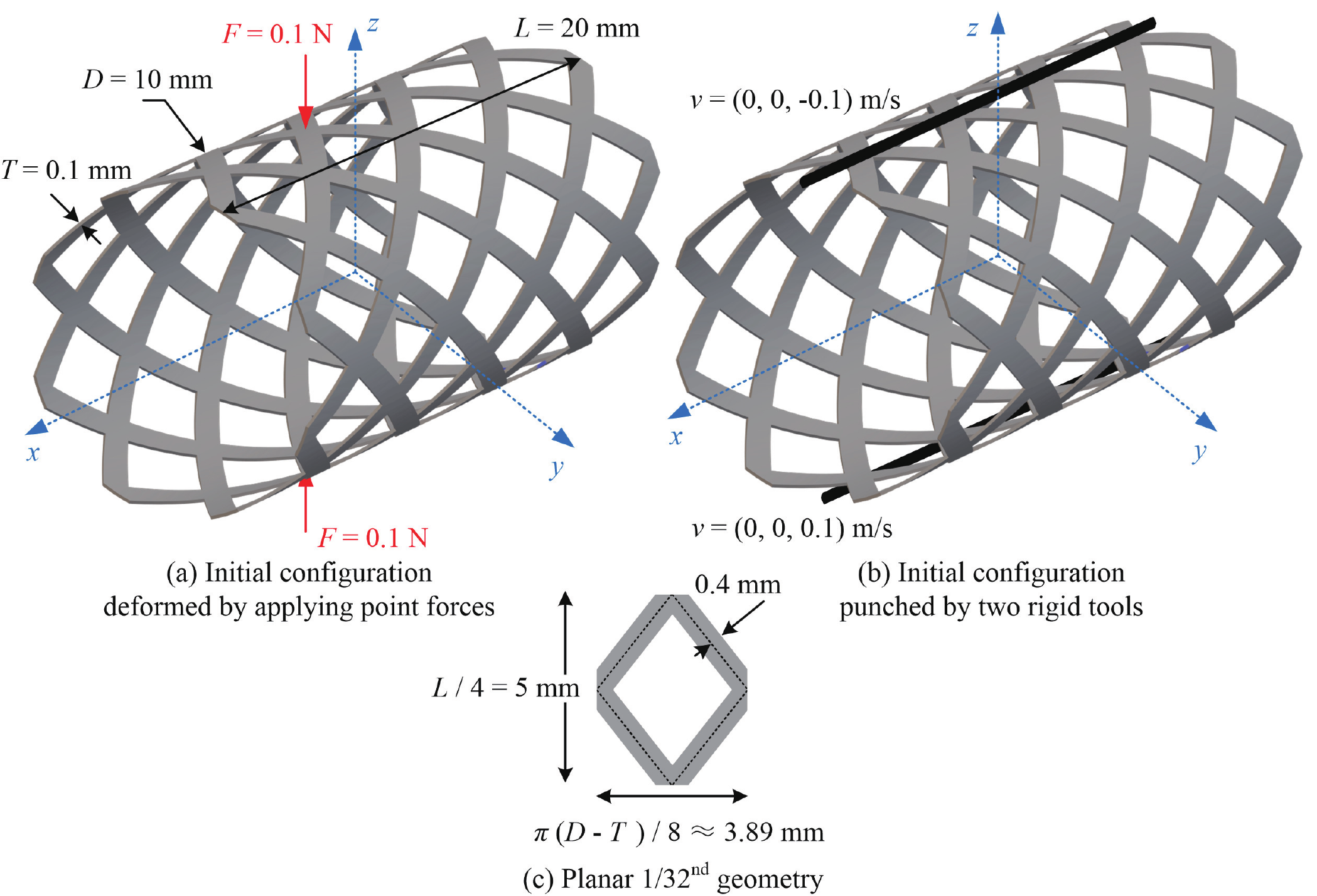}
	\caption{Stent structure: Problem setup. The corresponding computer-aided design (CAD) file in STL format can be downloaded from our code repository or GrabCAD.}
	\label{figs:stent_setup}
\end{figure}

A realistic cardiovascular stent, widely used in biomedical applications,
is investigated in this section to demonstrate the robustness and versatility of the TL-SPH-HF.
As shown in Figure \ref{figs:stent_setup},
a Palmaz-Schatz shaped stent with the length of $L = 20~\text{mm}$,
outer diameter $D = 10~\text{mm}$ and thickness $T = 0.1~\text{mm}$ is considered herein.
One of the element structures on the planar surface is also shown on the bottom panel of Figure \ref{figs:stent_setup}.
The material properties are $\rho_0 = 1100 ~\text{kg} / \text{m}^3$, 
Young’s modulus $E = 17~\text{MPa}$ and Poisson’s ratio $\nu = 0.45$.
Also, the initial particle distribution is generated by the body-fitted particle generator \cite{zhu2021cad} with initial particle spacing $dp = T / 3$.
Two diametrically opposed point forces $F = 0.1~\text{N}$ are applied on the stent to active the deformation as shown in Figure \ref{figs:stent_setup}(a), 
and the stent is also punched by two rigid tools modeled as cuboids with dimensions 20 $\times$ 0.15 $\times$ 0.15 $\text{mm}^3$ with the punch velocity of $0.1~\text{m/s}$ as shown in Figure \ref{figs:stent_setup}(b) to further examine the robustness of the present formulation. 

\begin{figure}[htb!]
	\centering
	\includegraphics[trim = 2mm 6mm 2mm 2mm, width=\textwidth]{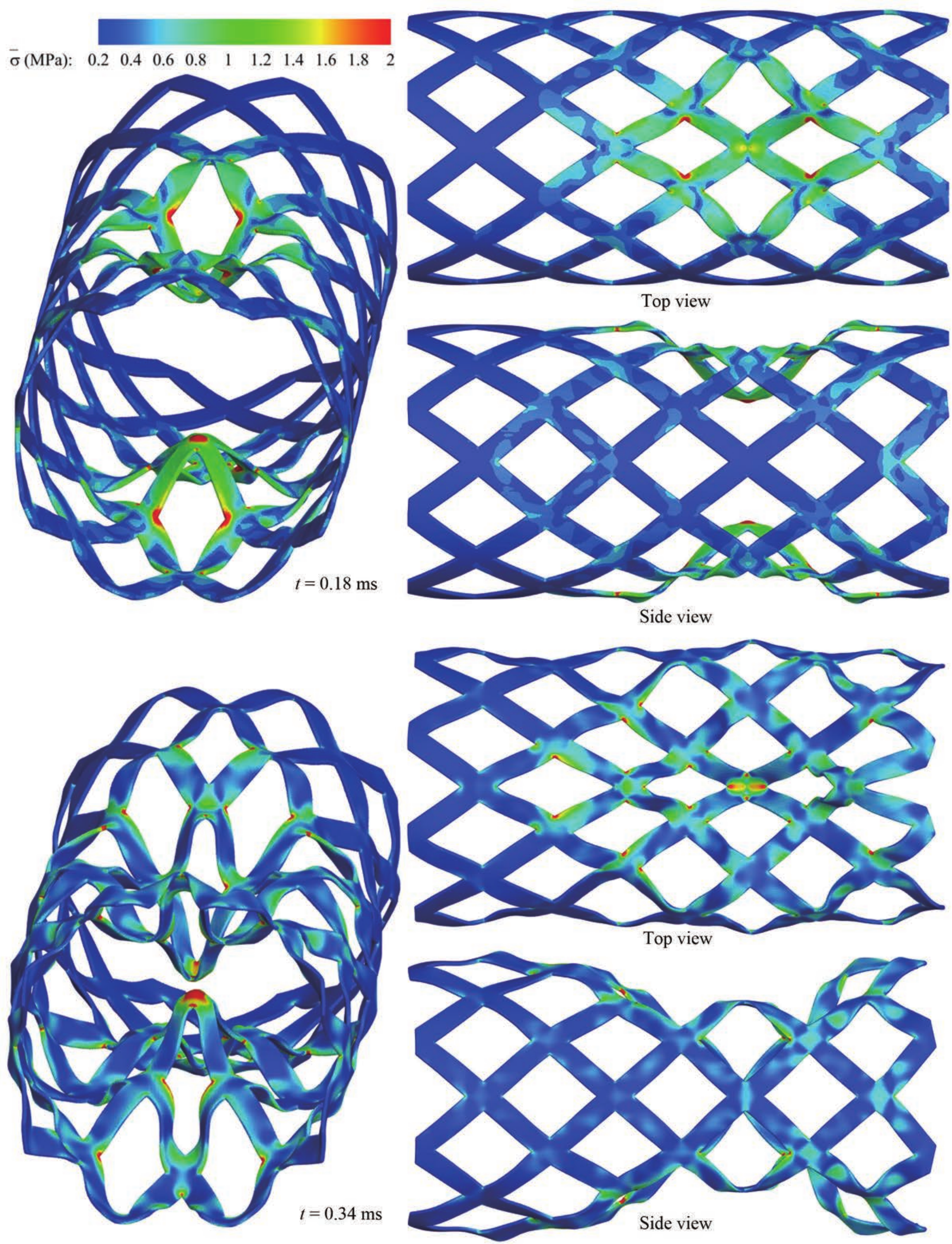}
	\caption{Stent structure: Deformed configuration under two diametrically opposed point forces $F = 0.1~\text{N}$ colored by von Mises stress at $t = 0.18~\text{ms}$ and $t = 0.34~\text{ms}$. The neo-Hookean material is applied with density $\rho_0 = 1100 ~\text{kg} / \text{m}^3$,  Young’s modulus $E = 17 ~\text{MPa}$ and Poisson’s ratio $\nu = 0.45$. }
	\label{figs:stent_point_force}
\end{figure}

\begin{figure}[htb!]
	\centering
	\includegraphics[trim = 2mm 6mm 2mm 2mm, width=\textwidth]{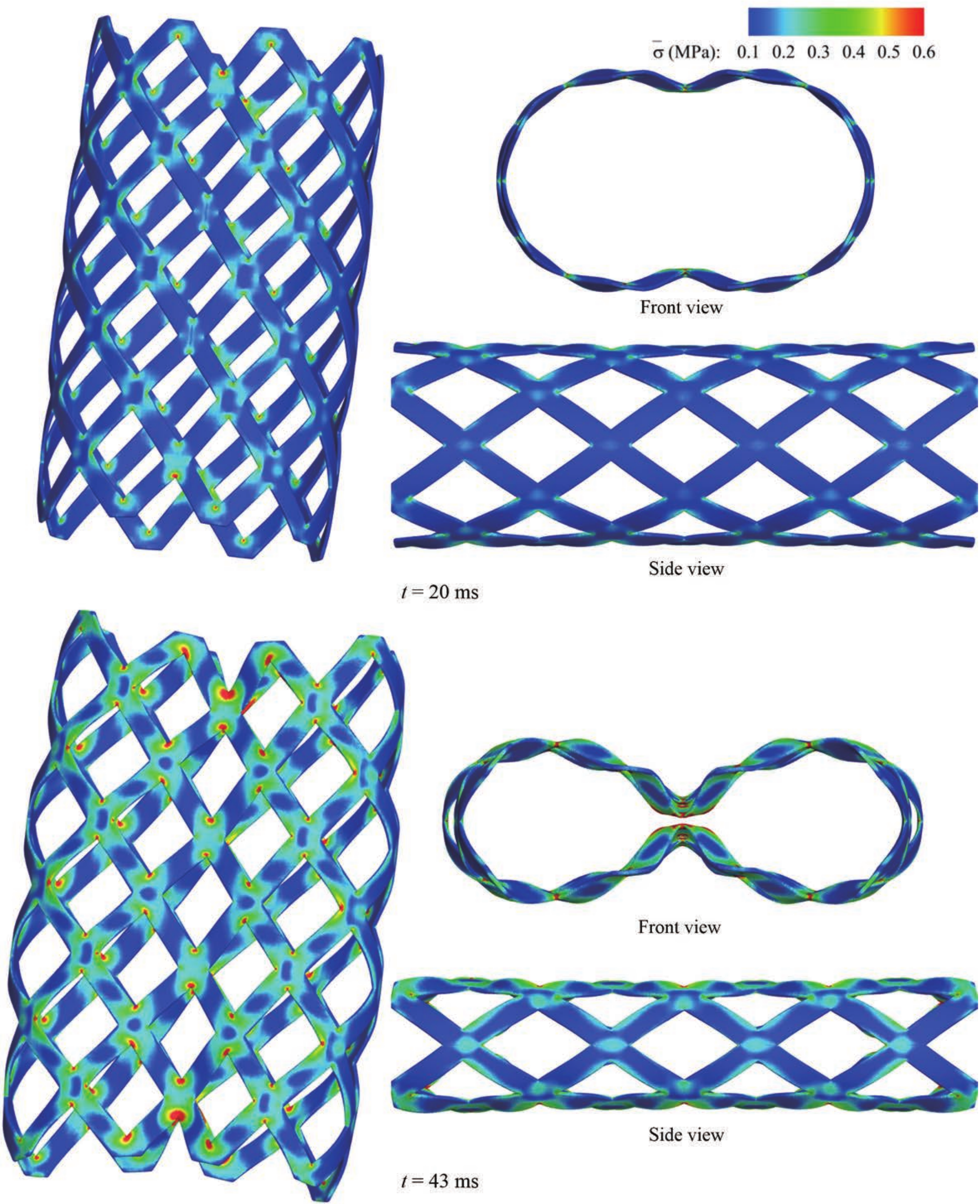}
	\caption{Stent structure: Deformed configuration under punching setup colored by von Mises stress at $t = 0.18~\text{ms}$ and $t = 0.34~\text{ms}$. 
		The neo-Hookean material is applied with density $\rho_0 = 1100 ~\text{kg} / \text{m}^3$,  Young’s modulus $E = 17 ~\text{MPa}$ and Poisson’s ratio $\nu = 0.45$. }
	\label{figs:stent_punching}
\end{figure}

Figure \ref{figs:stent_point_force} shows the overall deformation of the stent structure under point forces at time instants $t = 0.18~\text{ms}$ and $t = 0.34~\text{ms}$ with the von Mises stress contour.
The deformation pattern and smooth stress field of this complex thin structure are well captured, 
especially around the points of applying forces and sharp corners of the stent where the maximum stress exists.
Figure \ref{figs:stent_punching} shows the compressed stent colored by von Mises stress at different time instants.
It is remarkable that the extremely large deformation is well captured 
and paving the way of realistic cardiovascular applications.

\section{Concluding remarks}\label{sec:conclusion}
In this paper, 
we present an efficient, robust and hourglass-free formulation 
without introducing case-dependent tuning parameter and 
extra computational effort for the TL-SPH method. 
The proposed formulation demonstrates its capability of suppressing the long standing issues of the hourglass modes 
and shows its robustness in the simulation large strain dynamics. 
Last but not least, 
the deformation of complex stent structures is studied to demonstrate the versatility of the presented formulation, 
representing a stepping stone to practical applications in the field of biomechanics.
%
%
\section*{CRediT authorship contribution statement}
{\bfseries  D. Wu:} Conceptualization, Methodology, Investigation, Visualization, Validation, Formal analysis, Writing - original draft, Writing - review and editing; 
{\bfseries  C. Zhang:} Investigation, Methodology, Formal analysis, Writing - review and editing;
{\bfseries  X.J. Tang:} Investigation, Writing - review and editing;
{\bfseries  X.Y, Hu:} Supervision, Methodology, Investigation, Writing - review and editing.
%
%
\section*{Declaration of competing interest }
The authors declare that they have no known competing financial interests 
or personal relationships that could have appeared to influence the work reported in this paper.
%
%
\section*{Acknowledgments}
D. Wu is partially supported by the China Scholarship Council (No. 201906130189). 
C. Zhang and X.Y. Hu would like to express their gratitude to the German Research Foundation (DFG) 
for their sponsorship of this research under grant number DFG HU1527/12-4.
\clearpage

\section*{References}
\vspace{-0.8cm}
\renewcommand{\refname}{}
\bibliographystyle{elsarticle-num}
\bibliography{IEEEabrv,mybibfile}

\begin{thebibliography}{10}
\expandafter\ifx\csname url\endcsname\relax
  \def\url#1{\texttt{#1}}\fi
\expandafter\ifx\csname urlprefix\endcsname\relax\def\urlprefix{URL }\fi
\expandafter\ifx\csname href\endcsname\relax
  \def\href#1#2{#2} \def\path#1{#1}\fi

\bibitem{lucy1977numerical}
L.~B. Lucy, {A numerical approach to the testing of the fission hypothesis},
  The Astronomical Journal 82 (1977) 1013--1024.

\bibitem{gingold1977smoothed}
R.~A. Gingold, J.~J. Monaghan, {Smoothed particle hydrodynamics: theory and
  application to non-spherical stars}, Monthly Notices of the Royal
  Astronomical Society 181~(3) (1977) 375--389.

\bibitem{randles1996smoothed}
P.~Randles, L.~D. Libersky, Smoothed particle hydrodynamics: some recent
  improvements and applications, Computer Methods in Applied Mechanics and
  Engineering 139~(1-4) (1996) 375--408.

\bibitem{luo2021particle}
M.~Luo, A.~Khayyer, P.~Lin, {Particle methods in ocean and coastal
  engineering}, Applied Ocean Research 114 (2021) 102734.

\bibitem{khayyer2022systematic}
A.~Khayyer, H.~Gotoh, Y.~Shimizu, {On systematic development of FSI solvers in
  the context of particle methods}, Journal of Hydrodynamics (2022) 1--13.

\bibitem{monaghan2005smoothed}
J.~J. Monaghan, {Smoothed particle hydrodynamics}, Reports on Progress in
  Physics 68~(8) (2005) 1703.

\bibitem{liu2010smoothed}
M.~Liu, G.~Liu, {Smoothed particle hydrodynamics (SPH): an overview and recent
  developments}, Archives of Computational Methods in Engineering 17~(1) (2010)
  25--76.

\bibitem{monaghan2012smoothed}
J.~J. Monaghan, {Smoothed particle hydrodynamics and its diverse applications},
  Annual Review of Fluid Mechanics 44 (2012) 323--346.

\bibitem{zhang2021sphinxsys}
C.~Zhang, M.~Rezavand, Y.~Zhu, Y.~Yu, D.~Wu, W.~Zhang, J.~Wang, X.~Hu,
  {SPH}in{X}sys: {A}n open-source multi-physics and multi-resolution library
  based on smoothed particle hydrodynamics, Computer Physics Communications
  (2021) 108066.

\bibitem{sun2021accurate}
P.-N. Sun, D.~Le~Touze, G.~Oger, A.-M. Zhang, {An accurate FSI-SPH modeling of
  challenging fluid-structure interaction problems in two and three
  dimensions}, Ocean Engineering 221 (2021) 108552.

\bibitem{matthies2003partitioned}
H.~G. Matthies, J.~Steindorf, Partitioned strong coupling algorithms for
  fluid--structure interaction, Computers \& structures 81~(8-11) (2003)
  805--812.

\bibitem{matthies2006algorithms}
H.~G. Matthies, R.~Niekamp, J.~Steindorf, Algorithms for strong coupling
  procedures, Computer methods in applied mechanics and engineering 195~(17-18)
  (2006) 2028--2049.

\bibitem{yang2012free}
Q.~Yang, V.~Jones, L.~McCue, {Free-surface flow interactions with deformable
  structures using an SPH--FEM model}, Ocean Engineering 55 (2012) 136--147.

\bibitem{hermange20193d}
C.~Hermange, G.~Oger, Y.~Le~Chenadec, D.~Le~Touz{\'e}, {A 3D SPH--FE coupling
  for FSI problems and its application to tire hydroplaning simulations on
  rough ground}, Computer Methods in Applied Mechanics and Engineering 355
  (2019) 558--590.

\bibitem{antoci2007numerical}
C.~Antoci, M.~Gallati, S.~Sibilla, {Numerical simulation of fluid--structure
  interaction by SPH}, Computers \& structures 85~(11-14) (2007) 879--890.

\bibitem{han2018sph}
L.~Han, X.~Hu, {SPH modeling of fluid-structure interaction}, Journal of
  Hydrodynamics 30~(1) (2018) 62--69.

\bibitem{liu2019smoothed}
M.~Liu, Z.~Zhang, {Smoothed particle hydrodynamics (SPH) for modeling
  fluid-structure interactions}, Science China Physics, Mechanics \& Astronomy
  62~(8) (2019) 1--38.

\bibitem{johnson1996sph}
G.~R. Johnson, R.~A. Stryk, S.~R. Beissel, {SPH for high velocity impact
  computations}, Computer Methods in Applied Mechanics and Engineering
  139~(1-4) (1996) 347--373.

\bibitem{vignjevic2006sph}
R.~Vignjevic, J.~R. Reveles, J.~Campbell, {SPH in a total Lagrangian
  formalism}, CMC-Tech Science Press- 4~(3) (2006) 181.

\bibitem{liu2006restoring}
M.~Liu, G.-R. Liu, Restoring particle consistency in smoothed particle
  hydrodynamics, Applied Numerical Mathematics 56~(1) (2006) 19--36.

\bibitem{swegle1995smoothed}
J.~W. Swegle, D.~L. Hicks, S.~W. Attaway, {Smoothed particle hydrodynamics
  stability analysis}, Journal of Computational Physics 116~(1) (1995)
  123--134.

\bibitem{lind2020review}
S.~J. Lind, B.~D. Rogers, P.~K. Stansby, {Review of smoothed particle
  hydrodynamics: towards converged Lagrangian flow modelling}, Proceedings of
  the Royal Society A 476~(2241) (2020) 20190801.

\bibitem{rabczuk2004stable}
T.~Rabczuk, T.~Belytschko, S.~Xiao, {Stable particle methods based on
  Lagrangian kernels}, Computer Methods in Applied Mechanics and Engineering
  193~(12-14) (2004) 1035--1063.

\bibitem{monaghan2000sph}
J.~J. Monaghan, Sph without a tensile instability, Journal of computational
  physics 159~(2) (2000) 290--311.

\bibitem{gray2001sph}
J.~P. Gray, J.~J. Monaghan, R.~Swift, {SPH elastic dynamics}, Computer Methods
  in Applied Mechanics and Engineering 190~(49-50) (2001) 6641--6662.

\bibitem{owen2004tensor}
J.~M. Owen, {A tensor artificial viscosity for SPH}, Journal of Computational
  Physics 201~(2) (2004) 601--629.

\bibitem{zhang2017generalized}
C.~Zhang, X.~Y. Hu, N.~A. Adams, A generalized transport-velocity formulation
  for smoothed particle hydrodynamics, Journal of Computational Physics 337
  (2017) 216--232.

\bibitem{zhu2021consistency}
Y.~Zhu, C.~Zhang, X.~Hu, {A consistency-driven particle-advection formulation
  for weakly-compressible smoothed particle hydrodynamics}, Computers \& Fluids
  230 (2021) 105140.

\bibitem{belytschko2000unified}
T.~Belytschko, Y.~Guo, W.~Kam~Liu, S.~Ping~Xiao, {A unified stability analysis
  of meshless particle methods}, International Journal for Numerical Methods in
  Engineering 48~(9) (2000) 1359--1400.

\bibitem{bonet2002alternative}
J.~Bonet, S.~Kulasegaram, {Alternative total Lagrangian formulations for
  corrected smooth particle hydrodynamics (CSPH) methods in large strain
  dynamic problems}, Revue Europ{\'e}enne des {\'E}l{\'e}ments Finis 11~(7-8)
  (2002) 893--912.

\bibitem{de2013total}
T.~De~Vuyst, R.~Vignjevic, {Total Lagrangian SPH modelling of necking and
  fracture in electromagnetically driven rings}, International Journal of
  Fracture 180~(1) (2013) 53--70.

\bibitem{ba2018thermomechanical}
K.~Ba, A.~Gakwaya, {Thermomechanical total Lagrangian SPH formulation for solid
  mechanics in large deformation problems}, Computer Methods in Applied
  Mechanics and Engineering 342 (2018) 458--473.

\bibitem{maurel2008sph}
B.~Maurel, A.~Combescure, {An SPH shell formulation for plasticity and fracture
  analysis in explicit dynamics}, International Journal for Numerical Methods
  in Engineering 76~(7) (2008) 949--971.

\bibitem{lin2014efficient}
J.~Lin, H.~Naceur, D.~Coutellier, A.~Laksimi, {Efficient meshless SPH method
  for the numerical modeling of thick shell structures undergoing large
  deformations}, International Journal of Non-Linear Mechanics 65 (2014) 1--13.

\bibitem{peng2018thick}
Y.~Peng, A.~Zhang, F.~Ming, {A thick shell model based on reproducing kernel
  particle method and its application in geometrically nonlinear analysis},
  Computational Mechanics 62~(3) (2018) 309--321.

\bibitem{khayyer2018enhanced}
A.~Khayyer, H.~Gotoh, H.~Falahaty, Y.~Shimizu, {An enhanced ISPH--SPH coupled
  method for simulation of incompressible fluid--elastic structure
  interactions}, Computer Physics Communications 232 (2018) 139--164.

\bibitem{zhang2021multi}
C.~Zhang, M.~Rezavand, X.~Hu, {A multi-resolution SPH method for
  fluid-structure interactions}, Journal of Computational Physics 429 (2021)
  110028.

\bibitem{zhang2021integrative}
C.~Zhang, J.~Wang, M.~Rezavand, D.~Wu, X.~Hu, An integrative smoothed particle
  hydrodynamics method for modeling cardiac function, Computer Methods in
  Applied Mechanics and Engineering 381 (2021) 113847.

\bibitem{flanagan1981uniform}
D.~Flanagan, T.~Belytschko, {A uniform strain hexahedron and quadrilateral with
  orthogonal hourglass control}, International Journal for Numerical Methods in
  Engineering 17~(5) (1981) 679--706.

\bibitem{jacquotte1984analysis}
O.-P. Jacquotte, J.~T. Oden, {Analysis of hourglass instabilities and control
  in underintegrated finite element methods}, Computer Methods in Applied
  Mechanics and Engineering 44~(3) (1984) 339--363.

\bibitem{dyka1997stress}
C.~Dyka, P.~Randles, R.~Ingel, {Stress points for tension instability in SPH},
  International Journal for Numerical Methods in Engineering 40~(13) (1997)
  2325--2341.

\bibitem{vignjevic2000treatment}
R.~Vignjevic, J.~Campbell, L.~Libersky, A treatment of zero-energy modes in the
  smoothed particle hydrodynamics method, Computer Methods in Applied Mechanics
  and Engineering 184~(1) (2000) 67--85.

\bibitem{vignjevic2009review}
R.~Vignjevic, J.~Campbell, {Review of development of the smooth particle
  hydrodynamics (SPH) method}, in: Predictive Modeling of Dynamic Processes,
  Springer, 2009, pp. 367--396.

\bibitem{beissel1996nodal}
S.~Beissel, T.~Belytschko, {Nodal integration of the element-free Galerkin
  method}, Computer Methods in Applied Mechanics and Engineering 139~(1-4)
  (1996) 49--74.

\bibitem{vidal2007stabilized}
Y.~Vidal, J.~Bonet, A.~Huerta, {Stabilized updated Lagrangian corrected SPH for
  explicit dynamic problems}, International Journal for Numerical Methods in
  Engineering 69~(13) (2007) 2687--2710.

\bibitem{o2021fluid}
J.~O’Connor, B.~D. Rogers, {A fluid--structure interaction model for
  free-surface flows and flexible structures using smoothed particle
  hydrodynamics on a GPU}, Journal of Fluids and Structures 104 (2021) 103312.

\bibitem{randles2000normalized}
P.~Randles, L.~Libersky, {Normalized SPH with stress points}, International
  Journal for Numerical Methods in Engineering 48~(10) (2000) 1445--1462.

\bibitem{islam2019stabilized}
M.~R.~I. Islam, C.~Peng, {A stabilized total-Lagrangian SPH method for large
  deformation and failure in geomaterials}, arXiv preprint arXiv:1907.06990
  (2019).

\bibitem{lee2016new}
C.~H. Lee, A.~J. Gil, G.~Greto, S.~Kulasegaram, J.~Bonet, {A new
  Jameson--Schmidt--Turkel smooth particle hydrodynamics algorithm for large
  strain explicit fast dynamics}, Computer Methods in Applied Mechanics and
  Engineering 311 (2016) 71--111.

\bibitem{zhang2022artificial}
C.~Zhang, Y.~Zhu, Y.~Yu, D.~Wu, M.~Rezavand, S.~Shao, X.~Hu, {An artificial
  damping method for total Lagrangian SPH method with application in
  biomechanics}, Engineering Analysis with Boundary Elements 143 (2022) 1--13.

\bibitem{ganzenmuller2015hourglass}
G.~C. Ganzenm{\"u}ller, {An hourglass control algorithm for Lagrangian smooth
  particle hydrodynamics}, Computer Methods in Applied Mechanics and
  Engineering 286 (2015) 87--106.

\bibitem{belytschko1983correction}
T.~Belytschko, {Correction of article by DP Flanagan and T. Belytschko},
  International Journal for Numerical Methods in Engineering 19~(3) (1983)
  467--468.

\bibitem{stainier1994improved}
L.~Stainier, J.~P. Ponthot, {An improved one-point integration method for large
  strain elastoplastic analysis}, Computer Methods in Applied Mechanics and
  Engineering 118~(1-2) (1994) 163--177.

\bibitem{morris1997modeling}
J.~P. Morris, P.~J. Fox, Y.~Zhu, {Modeling low Reynolds number incompressible
  flows using SPH}, Journal of Computational Physics 136~(1) (1997) 214--226.

\bibitem{hu2006multi}
X.~Y. Hu, N.~A. Adams, A multi-phase sph method for macroscopic and mesoscopic
  flows, Journal of Computational Physics 213~(2) (2006) 844--861.

\bibitem{simo2006computational}
J.~C. Simo, T.~J. Hughes, {Computational inelasticity}, Vol.~7, Springer
  Science \& Business Media, 2006.

\bibitem{zhang2020sphinxsys}
C.~Zhang, M.~Rezavand, Y.~Zhu, Y.~Yu, D.~Wu, W.~Zhang, S.~Zhang, J.~Wang,
  X.~Hu, {SPH}in{X}sys: {A}n open-source meshless, multi-resolution and
  multi-physics library, Software Impacts 6 (2020) 100033.

\bibitem{ogden1997non}
R.~W. Ogden, Non-linear elastic deformations, Courier Corporation, 1997.

\bibitem{yue2015continuum}
Y.~Yue, B.~Smith, C.~Batty, C.~Zheng, E.~Grinspun, {Continuum foam: A material
  point method for shear-dependent flows}, ACM Transactions on Graphics (TOG)
  34~(5) (2015) 1--20.

\bibitem{zhang2022review}
C.~Zhang, Y.~Zhu, D.~Wu, X.~Hu, {Review on Smoothed Particle Hydrodynamics:
  Methodology development and recent achievement}, arXiv preprint
  arXiv:2205.03074 (2022).

\bibitem{bonet2002simplified}
J.~Bonet, S.~Kulasegaram, A simplified approach to enhance the performance of
  smooth particle hydrodynamics methods, Applied Mathematics and Computation
  126~(2-3) (2002) 133--155.

\bibitem{takeda1994numerical}
H.~Takeda, S.~M. Miyama, M.~Sekiya, {Numerical simulation of viscous flow by
  smoothed particle hydrodynamics}, Progress of Theoretical Physics 92~(5)
  (1994) 939--960.

\bibitem{hu2006angular}
X.~Hu, N.~Adams, Angular-momentum conservative smoothed particle dynamics for
  incompressible viscous flows, Physics of Fluids 18~(10) (2006) 101702.

\bibitem{wendland1995piecewise}
H.~Wendland, Piecewise polynomial, positive definite and compactly supported
  radial functions of minimal degree, Adv. Comput. Math. 4~(1) (1995) 389--396.

\bibitem{landau1986course}
L.~D. Landau, E.~M. Lifchits, Course of theoretical physics: Theory of
  elasticity (1986).

\bibitem{chen1996reproducing}
J.~S. Chen, C.~Pan, C.~T. Wu, W.~K. Liu, {Reproducing kernel particle methods
  for large deformation analysis of non-linear structures}, Computer Methods in
  Applied Mechanics and Engineering 139~(1-4) (1996) 195--227.

\bibitem{chen1996pressure}
J.~S. Chen, C.~T. Wu, C.~Pan, {A pressure projection method for nearly
  incompressible rubber hyperelasticity, part II: Applications}, Journal of
  Applied Mechanics 63 (1996) 869--876.

\bibitem{zhu2022dynamic}
Y.~Zhu, C.~Zhang, X.~Hu, {A dynamic relaxation method with operator splitting
  and random-choice strategy for SPH}, Journal of Computational Physics (2022)
  111105.

\bibitem{greaves2011poisson}
G.~N. Greaves, A.~L. Greer, R.~S. Lakes, T.~Rouxel, Poisson's ratio and modern
  materials, Nature materials 10~(11) (2011) 823--837.

\bibitem{smith2018stable}
B.~Smith, F.~D. Goes, T.~Kim, Stable neo-hookean flesh simulation, ACM
  Transactions on Graphics (TOG) 37~(2) (2018) 1--15.

\bibitem{zhu2021cad}
Y.~Zhu, C.~Zhang, Y.~Yu, X.~Hu, {A CAD-compatible body-fitted particle
  generator for arbitrarily complex geometry and its application to
  wave-structure interaction}, Journal of Hydrodynamics 33~(2) (2021) 195--206.

\bibitem{aguirre2014vertex}
M.~Aguirre, A.~J. Gil, J.~Bonet, A.~A. Carre{\~n}o, {A vertex centred finite
  volume Jameson--Schmidt--Turkel (JST) algorithm for a mixed conservation
  formulation in solid dynamics}, Journal of Computational Physics 259 (2014)
  672--699.

\bibitem{lee2019total}
C.~H. Lee, A.~J. Gil, A.~Ghavamian, J.~Bonet, {A total Lagrangian upwind smooth
  particle hydrodynamics algorithm for large strain explicit solid dynamics},
  Computer Methods in Applied Mechanics and Engineering 344 (2019) 209--250.

\end{thebibliography}

%
%
\end{document}